\documentclass[11pt]{article}

\usepackage[margin=1in]{geometry}
\usepackage{authblk}
\usepackage{graphicx}

\usepackage{natbib}
\usepackage{amsmath}
\DeclareMathOperator{\Ein}{Ein}
\DeclareMathOperator{\Lambert}{{\cal W}}

\newcommand{\target}[1]{\hat{m}_{#1}}

\begin{document}

\title{Cross-situational learning of large lexicons with finite memory}
\author{James Holehouse}
\author{Richard A. Blythe\thanks{Corresponding author, R.A.Blythe@ed.ac.uk}}
\affil{SUPA, School of Physics and Astronomy, University of Edinburgh,\\ Peter Guthrie Tait Road, Edinburgh EH9 3FD, UK}
\date{\today}

\maketitle

\begin{abstract}
Cross-situational word learning, wherein a learner combines information about possible meanings of a word across multiple exposures, has previously been shown to be a very powerful strategy to acquire a large lexicon in a short time. However, this success may derive from idealizations that are made when modeling the word-learning process. In particular, an earlier model assumed that a learner could perfectly recall all previous instances of a word's use and the inferences that were drawn about its meaning. In this work, we relax this assumption and determine the performance of a model cross-situational learner who forgets word-meaning associations over time. Our main finding is that it is possible for this learner to acquire a human-scale lexicon by adulthood with word-exposure and memory-decay rates that are consistent with empirical research on childhood word learning, as long as the degree of referential uncertainty is not too high or the learner employs a mutual exclusivity constraint. Our findings therefore suggest that successful word learning does not necessarily demand either highly accurate long-term tracking of word and meaning statistics or hypothesis-testing strategies.
\end{abstract}

\section{Introduction}

One of the many complex problems that needs to be solved during childhood language acquisition is the assignment of meanings to words. A major source of difficulty is referential uncertainty---that is, when an unfamiliar word is presented to a child, there are many possible meanings that the context of use might suggest, and the learner cannot be sure which meaning was intended \citep{Quine1960}.

Previous research suggests the following working hypothesis for how referential uncertainty is handled. First, in specific instances of a word's use, the learner is able to apply various heuristics that allow the space of likely meanings for the word to be reduced to a manageable number. Such heuristics include a bias towards assuming that words refer to whole objects rather than parts \citep{Macnamara1972}, an ability to infer the what a speaker is attending to \citep{Tomasello1986}, gaze direction \citep{Baldwin1991}, prior knowledge of language structure \citep{Gillette1999} or a mutual-exclusivity constraint \citep{Markman1988} that favors distinct meanings for different words. Acting in concert, these heuristics may combine to eliminate all referential uncertainty, delivering a single candidate meaning for the word, and moreover one that has a high probability of being the speaker's intended meaning. Such words are said to be \emph{fast mapped} \citep{Carey1978}.

It is unlikely that fast mapping can be achieved every time an unfamiliar word is presented to a child. Under these circumstances, the logical possibilities are to wait until fast mapping is possible, or to reduce the uncertainty that remains after application of heuristics by combining information from multiple such instances of a word's use. This latter process is referred to as \emph{cross-situational learning}, and both adults \citep[e.g.][]{Yu2007,Smith2011} and children \citep[e.g.][]{Smith2008,Scott2012,Suanda2014} have displayed this capability in an experimental setting. It is however possible that humans adopt a cross-situational strategy \emph{only} for the purpose of solving artificial word-learning tasks in the laboratory. \cite{Kachergis2014} have established that although this may be true to some extent, in that participants only reliably formed correct word-meaning relationships when explicitly asked to attend to co-occurring words and meanings, they nevertheless tracked co-occurrence statistics even when asked not to. This suggests that some degree of cross-situational learning is an automatic process, and potentially therefore one that would be utilized in a natural setting.

An unavoidable limitation of almost all word-learning experiments is that time constraints restrict the number of words and referents to a much smaller number than that encountered in the real world. Here, mathematical and computational models \citep{Siskind1996,Smith2001,Vogt2003,Smith2006} that encode specific strategies for word learning provide a useful tool for exploring their capability to acquire lexicons of a human scale \citep[i.e., around $60,000$ words,][]{Bloom2000}. At the heart of many models of cross-situational word learning is a measure of the strength of association between words and meanings that are known to the learner \citep[see e.g.][for a selection]{Smith2001,Vogt2003,Smith2006,Fazly2010,Tilles2012,Yu2012,Yurovsky2014,Rasanen2015,Hidaka2017}. The core dynamic in these models is that an association between a word and a meaning increases when a learner infers the latter as a potential referent of the former. The procedure for assigning a \emph{single} meaning to a word, given associations between that word and multiple meanings, is highly variable. One approach is for the learner to wait until one meaning is more strongly associated with a word than any other, at which point this becomes the word's meaning for that learner. If it is assumed that the target meaning is always correctly inferred as a possible meaning on each occasion of use (alongside other potential referents, which appear less consistently), this rule is equivalent to a process of elimination of the type discussed by \cite{Siskind1996}. Specifically, a learner initially assumes that a word could have any meaning, and narrows down the set of meanings according to the inferences drawn in each episode of a word's use. Once this set comprises a single meaning, the word is deemed learnt.

This specific eliminative word-learning process turns out to be sufficiently simple that it is possible to derive mathematical formul\ae\ that specify the time to learn a lexicon as a function of its size and other parameters that encode the degree of referential uncertainty \citep{Smith2006,Blythe2010}. The reason why this is possible boils down to the fact that a single counter-example to a word's hypothesized meaning is sufficient to exclude the hypothesis permanently. The main result of these analyses is that under a wide range of assumptions on word and meaning distributions, a human-scale lexicon of 60,000 words can be learnt using this strategy with 18 years' worth of linguistic input, and often much less \citep{Blythe2010}. An exception is when a non-target meaning is inferred alongside the target meaning with a very high probability: this competition slows the learning process down considerably \citep{Vogt2012}. However, these difficulties can be overcome if a mutual exclusivity constraint allows this competitor meaning to be excluded, and in fact under certain conditions, mutual exclusivity is powerful enough to render many words in the lexicon fast-mappable, i.e., unambiguous on a first encounter \citep{Reisenauer2013}.

A weakness of these models is that they rely on learners being able to keep track of word-meaning co-occurrence statistics over potentially large numbers of possible meanings and long times, perhaps to a greater extent than humans can reasonably achieve \citep{Medina2011}. It is therefore possible that the efficacy of cross-situational learning has been overstated, and that less cognitively demanding strategies should be sought. These include hypothesis-testing strategies, dubbed \emph{guess-and-test} \citep{Smith2011} or \emph{propose-but-verify} \citep{Trueswell2013}, wherein learners adopt a concrete hypothesis of a word's meaning on its first exposure, and switch to a different hypothesis if it is found sufficiently incompatible with subsequent instances of use. It is perhaps the case that these are instances of the same underlying strategy which has the appearance of eliminative cross-situational learning when the complexity of the learning problem is low, and of hypothesis testing when it is high \citep{Smith2011,Yurovsky2015}.

Our aim in this work is to understand how the rapid word-learning abilities of the ideal eliminative cross-situational learner formalized by \cite{Smith2006} are degraded by a finite memory constraint, and whether the degradation is so severe that additional strategies are then mandatory for learning a human-scale lexicon.  Previous work \citep{Yurovsky2015,Ibbotson2018} has explored the effect of memory decay in computational models of cross-situational learning, establishing for example that forgetting erroneous word-meaning associations can lead to a better overall learning outcome \citep{Ibbotson2018}. Here, our focus is on obtaining mathematical predictions for the time take to acquire a human-scale lexicon (i.e., $60,000$ words), as these conditions become cumbersome to simulate when learning algorithms become more complex.

Specifically, we allow word-meaning associations to decay exponentially between word exposures, and delay the elimination of candidate meanings until sufficient evidence has accrued that they are unlikely to be the target meaning for the word. As one would naturally expect, the time required to learn a lexicon increases as the learner's memory capacity decreases. However, we find that if associations persist on the timescale of a few weeks, as empirical studies suggest may be reasonable \citep{Wojcik2013}, a lexicon of $60,000$ words can be acquired given the number of words experienced by a child over an 18-year period that is in the middle of the range reported in the literature \citep{Hart1995,Sperry2018}.  Although the model learner's facility to acquire such a large lexicon is somewhat diminished if one adopts instead an amount of word exposure from the lower end of the range, we find that this is counteracted if the learner further employs a mutual exclusivity constraint to rule out candidate meanings for a word on the basis of other learnt words. In other words, we find a wide range of conditions under which a learner with human-like memory limitations can acquire a $60,000$-word lexicon without necessarily resorting to hypothesis-testing heuristics.

This paper is organized as follows. We first set out the definition of the word-learning model with memory decay in Section~\ref{sec:mod}, and recapitulate the main steps in determining the time required to learn a lexicon of a given size. The crucial quantity that is required in these calculations is the rate at which a single candidate meaning for a word is eliminated. In Section~\ref{sec:oneword} we explain how this is obtained for a learner with a finite memory constraint, and validate the predictions for the general case of multiple words and meanings with Monte Carlo simulations in Section~\ref{sec:multi}. We then, in Section~\ref{sec:60k}, use our mathematical results to scale up to human-size lexicons which lie beyond the reach of simulations. We conclude in Section~\ref{sec:disco} with a discussion of implications for theories of word learning.

\section{Modeling cross-situational learning}
\label{sec:mod}

\subsection{Model definition}
\label{sec:moddef}

We first set out the way that information about word-meaning relationships is made available to the model learner. The lexicon to learn comprises $W$ words. The word that is indexed by the integer $w$ is presented (e.g., spoken) to the learner as a Poisson process with rate $\phi_w$. That is, the mean time between presentations of word $w$ is $1/\phi_w$, and the mean time between presentations of any two words is $\Delta t = 1/\sum_w \phi_w$. We will usually rank the words in order of decreasing frequency, $\phi_{w+1} \le \phi_{w}$, and normalize the frequencies so that $\Delta t=1$. Then, after a time $t$, on average $t$ words will have been presented to the learner.

Whenever word $w$ is presented to the learner, they infer the meaning indexed by the integer $m$ with probability $a_{wm}$. This inference is assumed to result from the application of word learning heuristics of the type described in the introduction. For simplicity we assume that each meaning is inferred independently of any other: that is, both $m$ and $m'$ are inferred with probability $a_{wm}a_{wm'}$ when word $w$ is presented. Inferences made at different times are also independent. Our key assumption is that the target meaning of the word, denoted $\target{w}$, is \emph{always} among the set of inferred meanings (i.e., $a_{w\target{w}}=1$). We discuss the expected effect of relaxing this assumption in Section~\ref{sec:disco}.  Note that the mean number of inferences drawn at exposures of word $w$ is $\sum_m a_{wm}$. If the set of all possible meanings is infinite, this number need not be finite: nevertheless, the target meaning can still be successfully identified under these circumestances \citep{Blythe2016}.

For each word-meaning pair, the learner maintains an association strength $A_{wm}$ that is initially zero. When word $w$ is presented, and the learner infers $m$ as a possible meaning for the word, $A_{wm}$ is increased by $1$. Between exposures to the word, the associations decay exponentially towards zero at rate $\kappa$. That is, if time $t$ falls just after one exposure to word $w$, and time $t'$ just before its next exposure, $A_{wm}(t') = A_{wm}(t) {\rm e}^{-\kappa (t'-t)}$ for all meanings $m$, including the target meaning. The memory decay rate $\kappa$ defines a memory lifetime $1/\kappa$ that is measured in units of individual word presentations. If a long time passes between presentations of word $w$ ($1/\phi_w \gg 1/\kappa$ or, equivalently, $\kappa \gg \phi_w$), the association strengths will be close to zero, i.e., close to the learner's initial state of knowledge about the lexicon.

We acknowledge at the outset than an exponential decay of association strength is not an assumption that has strong empirical support in the literature, and that a variety of forms for `forgetting curves' have been proposed. Alongside the exponential function, proposals include power-law decays \citep{Lewis2005,Yurovsky2015}, functions that asymptote at a finite value rather than decaying to zero \citep{Brown2003,Averell2011}, and an exponential decay with a rate that depends on the total number of exposures to a word-meaning pair \citep{Ibbotson2018}. It has even been questioned whether any simple law exists \citep{Roediger2008}. Despite this empirical uncertainty, we believe that the exponential function is nevertheless appropriate for the present purpose of understanding the effects of a finite memory constraint on cross-situational learning for a number of reasons. First, it allows us to generalize the formul\ae\ for learning times previously obtained for the infinite-memory case \citep{Smith2006,Blythe2010,Reisenauer2013} which is crucial in applications to lexicons of a human-scale (60,000 words) that lie beyond the reach of computer simulations. Second, the exponential function has a single, well-defined timescale, which can be straightforwardly compared with memory lifetimes quoted in the empirical literature. Finally, all of the alternative decay forms mentioned above, like the power law, correspond to learners with superior long-term retention of associations than those whose memories decay exponentially to zero. We would expect such learners in general to be able to acquire a lexicon more readily than those studied here: consequently the results we obtain can be regarded as a lower bound on the performance of a cross-situational learner with a short-term memory decay rate $\kappa$.

The definition of the word-learning model is completed by a specification of when a word has become learnt, i.e., the point at which a single well-defined meaning can be assigned to the word by the learner. This must derive in some way from the set of association strengths that is available to the learner at a given time. Here, we make a choice that generalizes the eliminative model formalized by \cite{Smith2006} in which a meaning $m$ is permanently excluded as a candidate meaning for word $w$ on the first occasion that $m$ is not inferred when $w$ is exposed to the learner. This process of elimination causes the set of candidate meanings to be reduced over time, and the word is learnt once only a single meaning remains in this set. By construction (i.e., the fact that the target meaning is always among the set of inferred meanings), this single remaining meaning must be the target meaning.

This eliminative process can be re-expressed in terms of association strengths as follows: a meaning $m$ is permanently excluded as a candidate meaning for word $w$ if there exists some other meaning $m'$ such that $A_{wm'} > A_{wm}$. To see this, consider the infinite memory case $\kappa=0$. Then the association strengths simply count the number of times that a word-meaning pair has been inferred. Any meaning $m$ that has been inferred fewer times than the target satisfies $A_{w\target{w}} > A_{wm}$ and will be eliminated, as required. 

With a finite memory, $\kappa>0$, it remains the case that $A_{w\target{w}} \ge A_{wm}$ for all non-target meanings $m$, with equality holding only for those meanings that have (like the target) been inferred every time that the word has been presented. Therefore, a finite memory does not affect the word learning process if candidate meanings are eliminated when the condition $A_{wm'} > A_{wm}$ holds. Consequently, if we are to retain the notion of eliminating candidate meanings over time, and want memory to have an effect, we are required to generalize this condition to include an \emph{elimination threshold} $\Gamma$. Specifically, the meaning $m$ is permanently excluded as a candidate meaning for word $w$ at the earliest instant that $A_{wm'} > A_{wm} + \Gamma$. Note that any $\Gamma<1$ is equivalent to the eliminative model of \cite{Smith2006}, since on the first occasion where word $w$ is exposed and meaning $m$ is not inferred, $A_{w\target{w}}$ will have increased by $1$, but $A_{wm}$ will not. The dynamics of the association strengths, and how the threshold criterion is implemented, is illustrated in Fig.~\ref{fig:assocdyn}. In Appendix~\ref{sec:algos}, we set out the steps of a Monte Carlo simulation algorithm that implement this model learner.

\begin{figure}
\begin{center}
\includegraphics[width=0.66\linewidth]{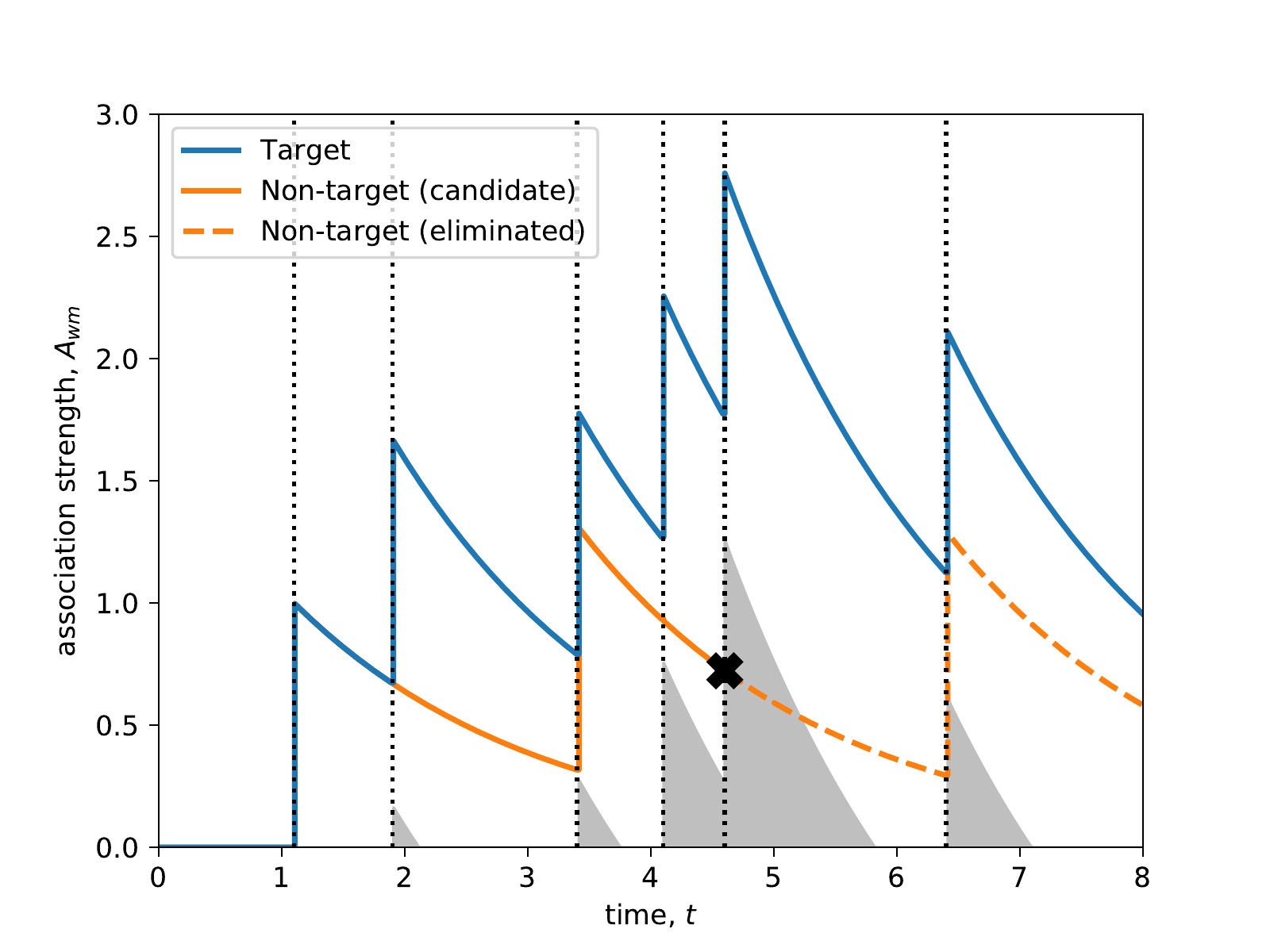}
\end{center}
\caption{\label{fig:assocdyn} Illustration of the association dynamics as a function of time for the target meaning (blue line) and a non-target meaning (orange line). The word is presented at the times indicated by the vertical dotted lines. At these times the association between the word and its target meaning increases by $1$; at a subset of these times the non-target meaning is also inferred and increases as well. Between word presentations, the association strengths decay exponentially. In this figure, the elimination threshold $\Gamma=1.5$: when the association between the word and the non-target meaning falls at least this far below that of the target meaning (indicated by the shaded region), the non-target meaning is permanently eliminated as a candidate for the word. This event is indicated by the cross.}
\end{figure}

It is worth pausing to establish a number of implications of this threshold for eliminating a candidate meaning. Most importantly, it retains the feature that the learner is guaranteed eventually to converge on the target meaning for each word. This is because the target meaning is always correctly inferred at each exposure of a word, and that once a candidate meaning has been excluded by crossing the threshold, it remains excluded forever. Starting from an initial condition in which the learner keeps an open mind as to the meaning of a word, the successive elimination of non-target meanings means that a concrete hypothesis for the word's meaning is made only once a single possible meaning remains. In this sense, one might regard the model of word learning as being conservative. A threshold $\Gamma>1$ is also conservative in the sense that an unambiguous instance of use (i.e., an occasion where precisely one meaning is inferred by the learner as a candidate for the word) is not sufficient for it to become learnt. The association between the word and its target meaning must be at least as large as $\Gamma$ for learning to occur, which means that the word must be encountered at least $\Gamma$ times before it is learnt. This means that this model precludes fast mapping (i.e., learning of a word on its first encounter), unless other mechanisms are brought into play. In Section~\ref{sec:mutex} we discuss one such mechanism, whereby a mutual exclusivity constraint allows a meaning to be eliminated immediately (i.e., without the threshold criterion being imposed). Another mechanism the provides for fast mapping is a propose-but-verify heuristic  (\citealp{Trueswell2013}; also referred to as ``guess-and-test'', \citealp{Blythe2010,Smith2011}), in which a concrete meaning is hypothesized for a word immediately after each exposure. The expected reduction of learning times that arise from this heuristic is described in Section~\ref{sec:disco}.

\subsection{Lexicon learning time}
\label{sec:tlearn}

The quantity of interest is the time taken for the learner to acquire the lexicon of $W$ words. In this section we review how this quantity is calculated, as this is foundational to the rest of this work.

As we are dealing with a stochastic model, the learning time is a random variable. In common with our earlier works \citep{Smith2006,Blythe2010,Reisenauer2013,Blythe2016} we define the learning time $t^\star$ through the probability $L(t)$ that the entire lexicon has been learnt by time $t$ via the implicit equation
\begin{equation}
\label{tstar}
L(t^\star) = 1 - \epsilon
\end{equation}
where $\epsilon$ is some small threshold. One way to interpret this equation is that if one takes a large population of learners, each exposed to a sequence of words drawn from the distribution defined in the previous section, then at time $t^\star$, a fraction $\epsilon$ of this population would not have learnt all the words in the lexicon.

It turns out that in many cases, calculating the learning time is fairly straightforward once one knows the rate $r_{wm}$ at which a non-target meaning $m$ is eliminated as a candidate for word $w$. Strictly speaking, this rate may change over time; however at late times (i.e., the time at which the lexicon is learnt to a high probability), the probability that $m$ is still entertained as a possible meaning for $w$ is given by ${\rm e}^{-r_{wm}t}$. If words are learnt independently---that is, if knowledge of one word's meaning does not allow inferences about to be drawn about any other word---then the probability that the lexicon has been learnt by time $t$ is then given by the probability that \emph{all} non-target meanings have been eliminated for \emph{all} words. That is,
\begin{equation}
\label{Lt}
L(t) = \prod_{w=1}^{W} \prod_{m \ne \target{w}} \left[ 1 - {\rm e}^{-r_{wm} t} \right] \;.
\end{equation}
The learning time is then determined by combining (\ref{tstar}) and (\ref{Lt}), and solving for $t^\star$. In practice, one often finds that it is sufficient to determine the \emph{slowest} elimination rates $r_{wm}$ to obtain a good estimate of $t^\star$. This simplifies the form of (\ref{Lt}) to the extent that analytical formul\ae\ for $t^\star$ can be obtained, the value of which is that they circumvent the need to obtain learning times from simulation.

\begin{figure}
\begin{center}
\includegraphics[width=0.66\linewidth]{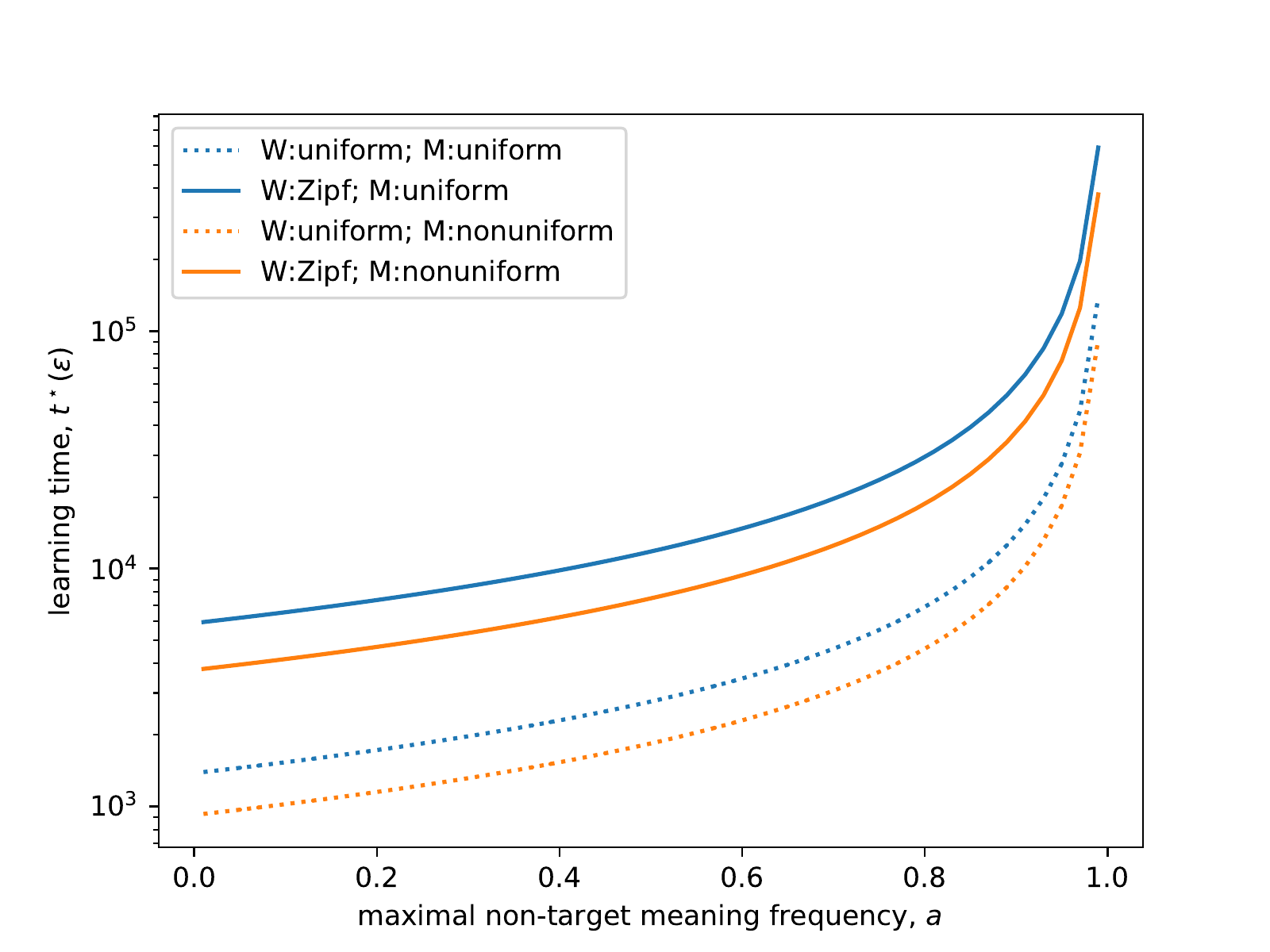}
\end{center}
\caption{\label{fig:classic} Learning times for a lexicon of $W=100$ words with $M=100$ non-target meanings against each with as a function of the frequency of the most commonly-inferred non-target meaning. Dotted and solid lines correspond to a uniform and Zipf word distribution, respectively. Blue and orange curves correspond to a uniform and nonuniform meaning distribution, respectively. The nonuniform case meaning distribution follows an inverse square root law, $a_{wm} = a / \sqrt{m}$. It is slightly easier to learn a lexicon with a nonuniform meaning distribution, because the lower frequency candidate meanings are rapidly eliminated.}
\end{figure}

We set out the general strategy for calculating the learning time in Appendix~\ref{sec:tcalc}. Here, we illustrate with the simplest case, where learners have infinite memory, the word frequency distribution is uniform, $\phi_i = \frac{1}{W}$, and where the frequency with which meanings are inferred is also uniform, $a_{wm}=a$.  To be clear, this means that when a learner encounters the word $w$, there are $M$ non-target meanings that the learner might infer as a candidate meaning, each with probability $a$. At any one instance of a word's use, the mean number of non-target meanings that is inferred as a possible meaning is $aM$. As discussed above, when the elimination threshold $\Gamma<1$, a candidate meaning is excluded the very first time that it fails to be inferred. Thus the rate at which a meaning is eliminated is equal to the rate at which it fails to be inferred in the context of a given word. If word $w$ is presented at rate $\phi_w$, this elimination rate is $r_{wm}=\phi_w(1-a)$. With a uniform word distribution $\phi_w=1/W$, and the elimination rate is the same for all $WM$ non-target meanings.  Thus here, one finds from (\ref{Lt}) that
\begin{equation}
L(t) = \left[1  -{\rm e}^{-r t} \right]^{WM}
\end{equation}
where $r=(1-a)/W$, the elimination rate common to all non-target meanings. Setting this equal to $1-\epsilon$, as in (\ref{tstar}) we find the learning time
\begin{equation}
\label{tuniform}
t^\star = -\frac{1}{r} \ln\left[ 1 - (1-\epsilon)^{\frac{1}{WM}} \right] \;.
\end{equation}
This result, plotted in Fig.~\ref{fig:classic}, is essentially the same as that given by \cite{Smith2006} and \cite{Blythe2010}. When we relax the assumption of infinite memory, below, it will still be the case that all non-target meanings are eliminated at a common rate $r$ when words and meanings both have uniform distributions, and thus a result of the form (\ref{tuniform}) will continue to apply. What changes is how the rate $r$ depends on the memory decay rate $\kappa$ and elimination threshold $\Gamma$, in a matter to be determined in Section~\ref{sec:oneword} below.

In reality, word forms do not have a uniform distribution, and it is highly unlikely that each non-target meaning has exactly the same probability of being inferred. It turns out that the easier case to handle is when the meaning distribution is non-uniform. In this case, we order the non-target meanings by the frequency with which they are inferred, from largest to smallest. If the difference in the inference frequencies of the two most common non-target meanings is sufficiently large, then at the time the most common meaning has been eliminated, it is very unlikely that any other meaning remains to be eliminated. The upshot of this is that for each word $w$, one needs to keep only the most frequent non-target meaning in (\ref{Lt}). This is equivalent to setting the total number of non-target meanings for each word, $M$, equal to $1$. If the maximum non-target inference frequency has the same value, $a$, for each word, the learning time is then given by (\ref{tuniform}) with $M=1$.  This also means that the lexicon is learnt more quickly than when the meanings are uniformly distributed (given the same frequency for the most commonly inferred non-target meaning), because the low-frequency meanings are rapidly eliminated---see Fig.~\ref{fig:classic}.

The case of a nonuniform word distribution is less straightforward to handle, and a detailed discussion is deferred to Appendix~\ref{sec:tcalc}. In this work, we restrict ourselves to the Zipf distribution
\begin{equation}
\label{zipf}
\phi_w = \frac{1}{Z} \frac{1}{w} \quad\mbox{where}\quad Z = \sum_{w=1}^{W} \frac{1}{w} \;.
\end{equation}
What makes this more complicated is that the learning time is determined by the words that are hardest to learn: these are the words in the tail of the frequency distribution ($w\approx W$), many of which have a similar frequency. This means that we cannot simply keep the least frequent word in (\ref{Lt}), but must instead retain a representative sample from the tail of the distribution. We set out the mathematical procedure that achieves this in Appendix~\ref{sec:tcalc}.

We consider first the case where each word $w$ is accompanied by $M$ non-target meanings, each of which has the same frequency $a$ of being inferred alongside the target meaning (i.e., the uniform meaning distribution as previously described). The elimination rate for a given meaning now depends on the word it is confounding, that is, $r_{wm} \equiv r_w = \phi_w (1-a)$.  The expression for the learning time in this case is
\begin{equation}
\label{tzipf}
t^\star = \frac{1}{r_W} \Lambert\left( \frac{r_W}{|r'_W|} \frac{M}{\epsilon} \right)
\end{equation}
where $\Lambert(z)$ is the principal branch of the Lambert W function \citep{Corless1996} and 
\begin{equation}
r'_W = \left. \frac{{\rm d}}{{\rm d} w} r_w \right|_{w=W} \;.
\end{equation}
In the case where the threshold $\Gamma<1$, $r_W = (1-a)/(WZ)$ and $r_W/|r'_W|=W$, and we recover previously published results \citep{Blythe2010}. In particular, we find that the learning time is approximately a factor $Z$ times longer when learning words that have a Zipf frequency distribution. When $W=100$, as in Fig.~\ref{fig:classic}, $Z\approx5.19$, and the learning times are systematically around $5$ times longer when words have a Zipf distribution than when they are uniform. A similar expression will again apply in the case of learning with a finite memory, with a suitably redefined set of elimination rates $r_w$. 

The generalization to the case of a nonuniform meaning distribution, also shown in Fig.~\ref{fig:classic}, is obtained in the same way (and under the same conditions) as it was for the case of a uniform word distribution.  Specifically, we keep only the most frequent non-target meaning for each word and set $M=1$ in (\ref{tzipf}). Recall that it is assumed for simplicity that this maximal frequency $a$ is the same for each word.

\section{Elimination rates under cross-situational learning with finite memory}
\label{sec:oneword}

In the previous section, we noted that the crucial quantity in predicting a lexicon learning time is the rate $r_{wm}$ at which the non-target meaning $m$ is eliminated as a candidate meaning for word $w$.  Here our aim is to determine this rate when associations between words and meanings decay at rate $\kappa$ as described in Section~\ref{sec:moddef}. For this purpose, it is sufficient to consider a single word, which is presented at rate $\phi$, and two meanings. One of these meanings, denoted $\target{w}$, is the target that is inferred at each exposure; the other, denoted $m$, is a non-target meaning that is inferred with probability $a$ at each exposure.

It is helpful to introduce the variable $x$ that measures the difference in the association strength between the target and non-target meaning, i.e., $x = A_{w\target{w}} - A_{wm}$. According the the dynamics defined in Section~\ref{sec:moddef}, whenever the word is presented with the non-target meaning present, both association strengths increase, so $x$ is unchanged. When the word is presented and the non-target meaning is not inferred, $x$ increases by one. This happens as a Poisson process with rate $\phi(1-a)$. Between word exposures, both association strengths decay exponentially at rate $\kappa$ towards zero. The difference between these association strengths therefore also decays towards zero at rate $\kappa$. The criterion for excluding the non-target meaning can then be stated as $x$ reaching the value $\Gamma$ for the first time.

To obtain an equation for the dynamics of $x$ that is amenable to numerical solution, it is convenient to recast the problem as one in which $x$ is quantized in steps of $1/N$, and $N$ is a suitably large number. The quantity $n$, defined via $x=n/N$, is then the number of steps away from the point at which the two meanings have the same association strength; the elimination threshold lies at $n=\Gamma N$. It is perhaps helpful to think of $n$ being analogous to the number of individuals in some population. The unit increase of the difference in association strengths then corresponds to $N$ new individuals being introduced to the population, an event that occurs at a rate $\phi(1-a)$. The exponential decay in the association difference can then be modelled by each individual in the population dying with a rate $\kappa$: the total rate at which an individual is removed from the population is then $n \kappa$.  Introducing $p_n(t)$ as the probability that there are $n$ individuals in the population at time $t$ (i.e., the association difference $x=n/N$) then satisfies the master equation
\begin{equation}
\label{master}
\frac{{\rm d}}{{\rm d} t} p_n(t) = \phi(1-a) \left[ p_{n-N}(t) - p_n(t) \right] + \kappa \left[ (n+1) p_{n+1}(t) - n p_n(t) \right] \;
\end{equation}

We now impose the boundary condition $p_{\Gamma N}(t)=0$ for all $t$, which amounts to the process being absorbed (stopping) when $x=\Gamma$, i.e., the threshold at which the non-target meaning is eliminated. This has the consequence that $p_n(t)$ is now to be interpreted as the \emph{joint} probability that the difference in association strengths is $n/N$ \emph{and} the non-target meaning has not been eliminated. At late times, this probability is expected to decay exponentially as $p_n(t) \sim \psi_n {\rm e}^{-r t}$, where $r$ is the elimination rate that we seek, and $\psi_n$ is an amplitude whose value does not enter into expressions for the learning time.  Substituting this ansatz into the master equation (\ref{master}), we obtain the eigenvalue equation
\begin{equation}
\label{eigen}
\psi_{n-N} - \psi_n + k \left[ (n+1) \psi_{n+1} - n \psi_n \right] = - \lambda(k) \psi_n
\end{equation}
where the elimination rate $r$ depends on the smallest eigenvalue $\lambda(k)$ through
\begin{equation}
\label{r1}
r = \phi(1-a) \lambda\left( \frac{\kappa}{\phi(1-a)} \right) \;.
\end{equation}

The simplest way to obtain $\lambda(k)$ is to solve the equation (\ref{eigen}) numerically, for which standard routines exist. In this work we use the \texttt{eigvals} routine in the NumPy linear algebra package \citep{Numpy}, and have found that $N=100$ provides a sufficiently fine discretization of the problem for our needs\footnote{All source code that performs the calculations and simulations presented in this work will be archived at a permanent URL on publication}. To convert the elimination rate $r$ to a learning time, we use Eq.~(\ref{tuniform}) with $W=M=1$, that is
\begin{equation}
\label{tstar1}
t^\star = - \frac{\ln \epsilon}{r} \;.
\end{equation}

\begin{figure}
\begin{center}
\includegraphics[width=0.66\linewidth]{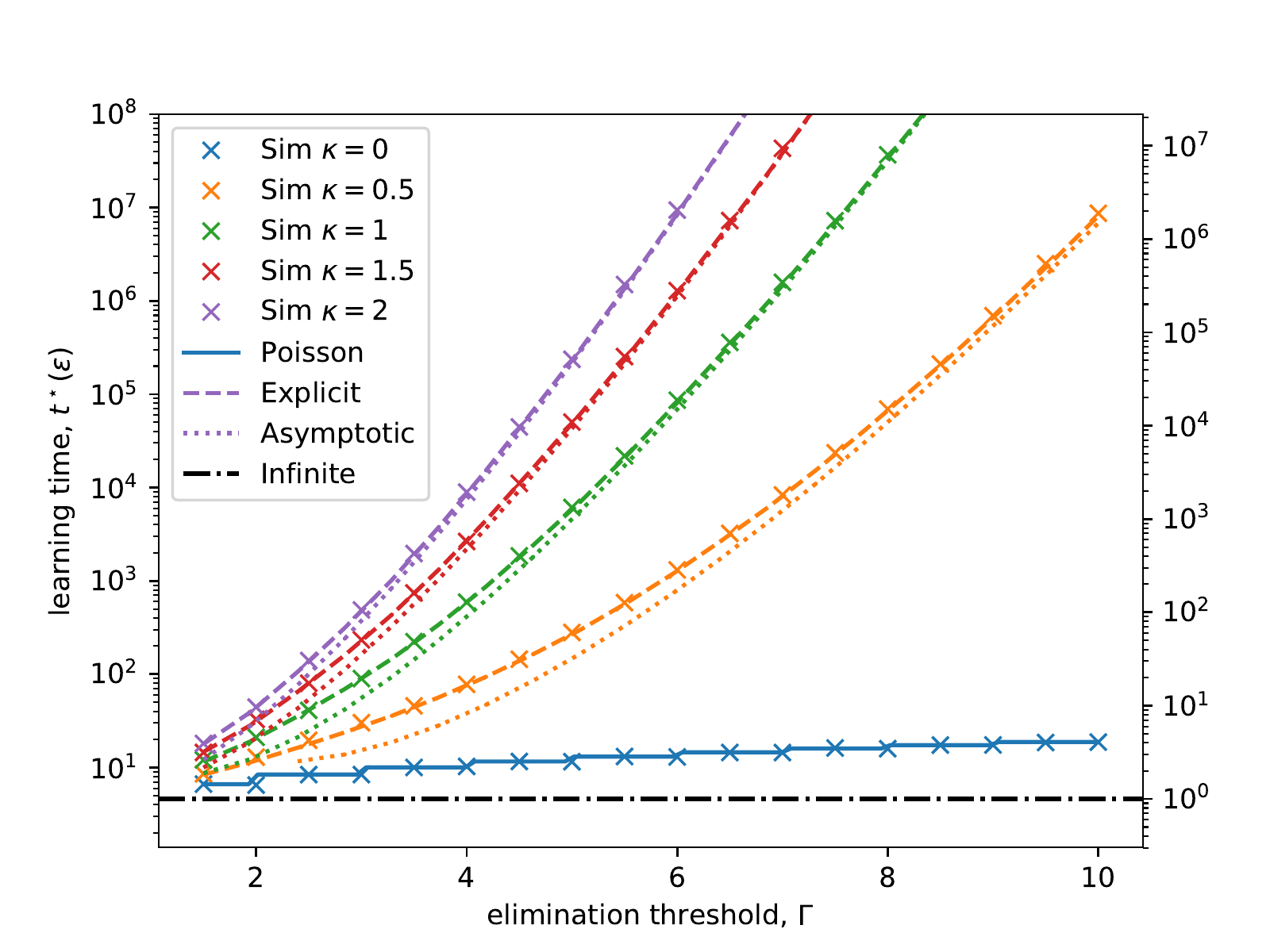}
\end{center}
\caption{\label{fig:oneword} Learning times for a single word with $\phi=1$, $a=0$, $\epsilon=0.01$ and various $\kappa$ as a function of the elimination threshold $\Gamma$. Crosses show learning times estimated from $10,000$ Monte Carlo samples of the word learning process (see Appendix~\ref{sec:algos}). The dashed lines show the learning time $t^\star$ given by Eqs.~(\ref{tstar1}) and (\ref{r1}), and where the eigenvalue $\lambda(k)$ has been calculated by explicit numerical solution of (\ref{eigen}). The dotted lines are obtained from the asymptotic expansion (\ref{lasymp}), valid in the limit of large $\kappa$ or large $\Gamma$. The solid line indicates the special case of infinite memory ($\kappa=0$), which is discussed in Appendix~\ref{sec:kappa0}. The chain indicates the learning time in the case of infinite memory and an elimination threshold $\Gamma<1$ previously considered \citep{Smith2006}. In this and subsequent plots, the right-hand vertical axis indicates the the time relative to this baseline (i.e., $10^2$ is $100$ times slower).}
\end{figure}

The validity of this result relies on other solutions of (\ref{eigen}) corresponding to more rapid decays that can be neglected at times of order $t^\star$. This we can test by comparing learning times predicted by the smallest eigenvalue and Eq.~(\ref{tstar1}) with direct Monte Carlo simulations of the word learning process (see Appendix~\ref{sec:algos} for an outline of the simulation algorithm). In these simulations we take $\phi=1$ and $a=0$ with no loss of generality; with these choices we have that  $k=\kappa$. In Fig.~\ref{fig:oneword} we see that this single solution of (\ref{eigen}) is sufficient to allow us to predict the learning time to good accuracy over a range of decay rates $\kappa$ and elimination thresholds $\Gamma$. We further see that the learning times can be extended by several orders of magnitude relative to previously studied cases \citep{Smith2006,Blythe2010} which correspond to $\Gamma<1$.

Recall that when the elimination threshold $\Gamma>1$, a learner with infinite memory ($\kappa=0$) will need to be exposed to a word multiple times for the non-target meaning to be eliminated, and therefore we expect the learning time to increase with the threshold $\Gamma$ in this case. This case is degenerate, and requires a separate mathematical treatment that is presented in Appendix~\ref{sec:kappa0}, with the result shown in Fig.~\ref{fig:oneword}. A key feature of the learning time in this regime is a step-wise increase in the learning time at integer values of the elimination threshold $\Gamma$. This arises from the fact that an extra exposure to the word is required in order for the non-target meaning to be discarded.

Ideally, one would like a general mathematical expression for the smallest eigenvalue of (\ref{eigen}), as this would allow one to understand more deeply how the elimination rate $r$ depends on the forgetting rate $\kappa$ and the elimination threshold $\lambda$. We have only been able to obtain such an expression in the limit of fast forgetting ($\kappa\to\infty$) or a large threshold ($\Gamma\to\infty$), both of which lead to extended learning times. Even then, both the calculation and the form of the final result are somewhat involved, and are therefore relegated to Appendix~\ref{sec:asymp}. Nevertheless, we see from Fig.~\ref{fig:oneword} that despite the assumption of rapid memory decay or a high threshold, the result gives a good prediction for the learning time over a wide range of parameter values. Since this formula does not involve numerical solution of the eigenvalue equation (\ref{eigen}), it may prove more convenient in certain applications.

For extremely large decay rates $\kappa$ or thresholds $\Gamma$, we obtain a limiting form for the learning time that is reasonably simple, and yields some insight. It reads
\begin{equation}
\label{bigbig}
t^\star \sim - \sqrt\frac{2\pi}{\Gamma} \frac{\ln\epsilon}{\kappa} \left(\frac{\kappa\Gamma}{\phi(1-a)}\right)^{\!\!\Gamma}
\end{equation}
which reveals that the learning time is much more sensitive to the elimination threshold $\Gamma$ than the memory decay $\kappa$. Specifically, $t^\star$ increases as a power of $\kappa$ and super-exponentially with $\Gamma$. This sensitivity to the threshold $\Gamma$ is an important feature of the model that we return to in Section~\ref{sec:disco} below.

\section{Validation of learning time predictions for small lexicons}
\label{sec:multi}

In Section~\ref{sec:tlearn}, we obtained mathematical predictions for the time taken to learn multiple words, each confounded by multiple non-target meanings, given the rates $r_{wm}$ that meaning $m$ is eliminated as a candidate for word $w$. In Section~\ref{sec:oneword}, we obtained an estimate of the rates $r_{wm}$ for  a cross-situational learner with a finite-memory constraint. In this section, we combine these results to obtain predictions for lexicons comprising multiple words and meanings, and confirm their validity by comparing to results from Monte Carlo simulations. This is key to scaling up to large (human-scale) lexicons in Section~\ref{sec:60k} below, where simulations become unfeasible and we are required to rely solely on the mathematical results (and the approximations that are inherent to them). The reader who is primarily interested in these latter results may wish to devote more attention to that section.

\subsection{Independent learning of words}

\begin{figure}
\begin{center}
\includegraphics[width=0.475\linewidth]{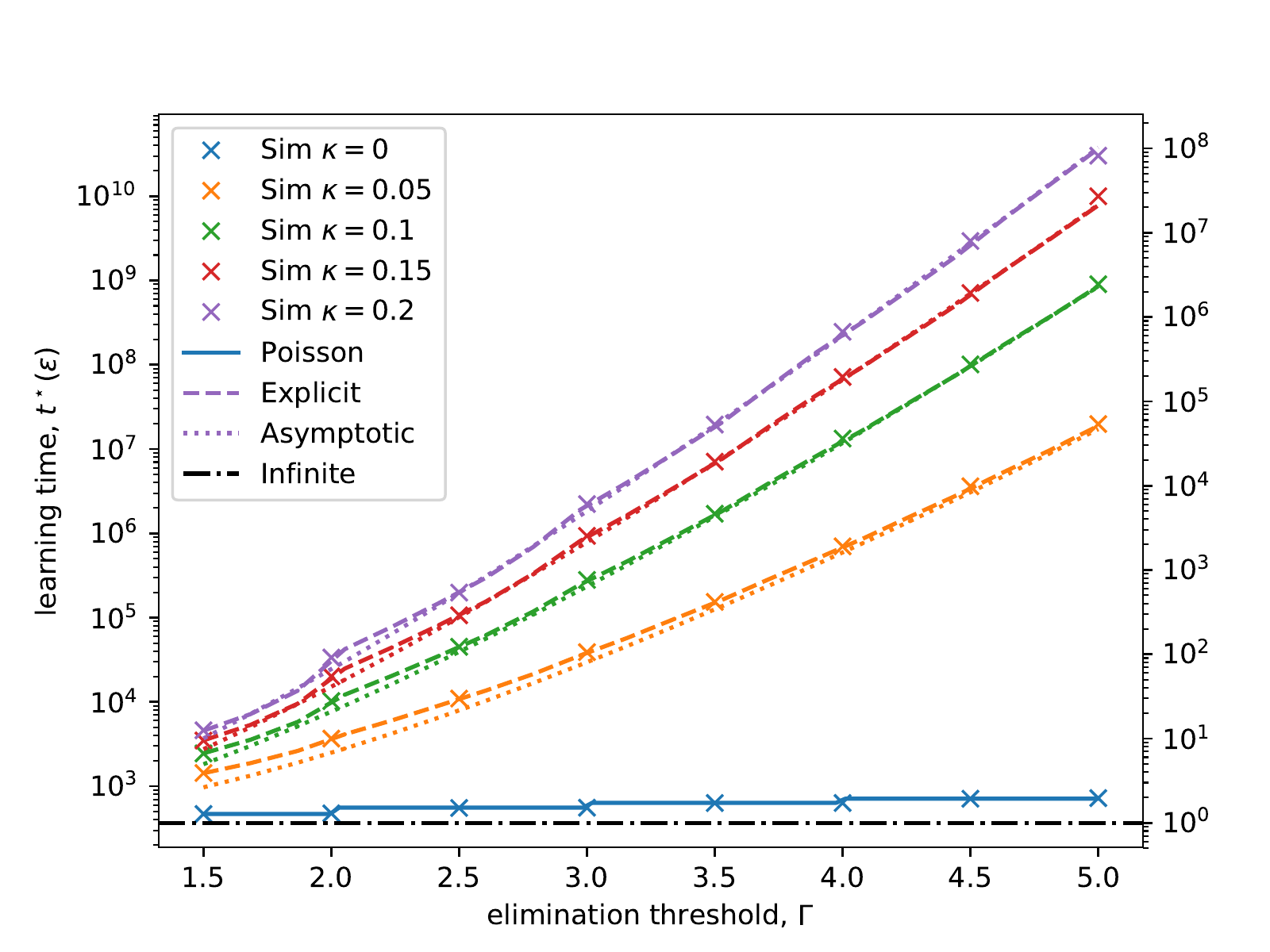}
\includegraphics[width=0.475\linewidth]{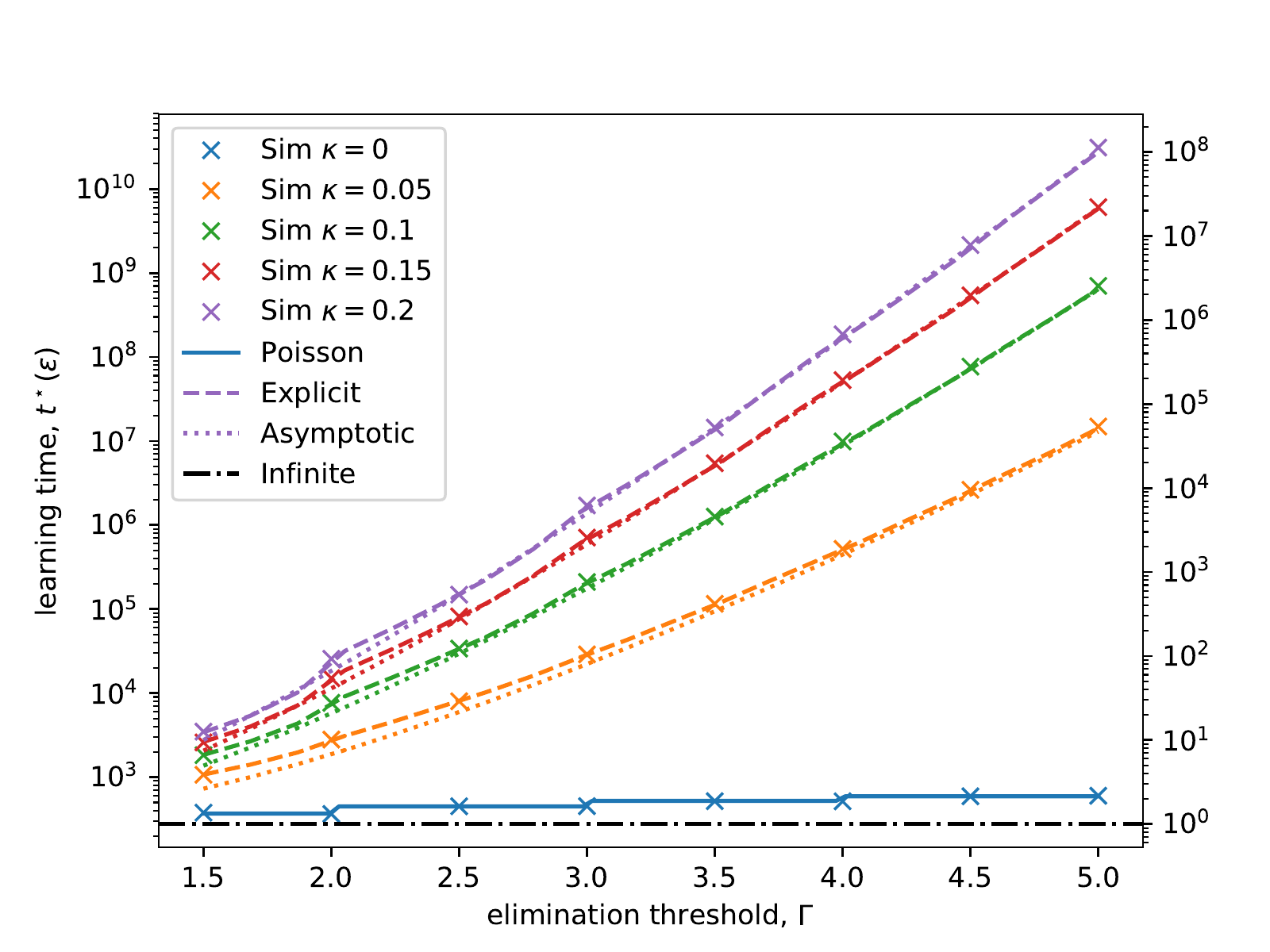}
\end{center}
\caption{\label{fig:w10m10uniform} Learning time as a function of the elimination threshold $\Gamma$ for a lexicon of $W=10$ words and $M=10$ non-target meanings for each. In both figures, the word frequency distribution is uniform. In the left-hand figure, the meaning frequency distribution is uniform, $a=\frac{3}{4}$, whilst in the right-hand figure it has the inverse square root form (\ref{invsqrt}), again with $a=\frac{3}{4}$.  As in Fig.~\ref{fig:oneword}, the threshold $\epsilon=0.01$, crosses indicate learning times obtained from Monte Carlo simulations, solid lines are exact results for $\kappa=0$ (see Appendix~\ref{sec:kappa0}), dashed lines are found by numerical computation of the eigenvalue (\ref{eigen}), and dotted lines via the asymptotic result (\ref{lasymp}). The horizontal chain indicates the learning time for a learner with infinite memory and an elimination threshold $\Gamma<1$. The size of the Monte Carlo samples depends on the learning time: for $t^\star$ up to around $10^7$, we are able to generate $10,000$ samples, but constraints on simulation runtimes reduce this to around $30$ in the most extreme cases.}
\end{figure}

Recall that we have considerable freedom in the choice of word and meaning frequency distributions. We begin with the simplest case, where both have a uniform distribution. As previously described in Section~\ref{sec:tlearn}, with $W$ words in the lexicon, each word $w$ is exposed at rate $\phi_w=1/W$. When a word is presented to the learner, each of a set of $M$ non-target meanings $m$ is independently inferred as a candidate meaning with probability $a$.  The learning time for this case is given by the expression (\ref{tuniform}), but with the elimination rate $r$ now given by the expression that applies when associations decay over time, i.e., (\ref{r1}).  We compare this analytical prediction with Monte Carlo simulation results in Fig.~\ref{fig:w10m10uniform}. In these simulations, we took $W=10$ words, $M=10$ non-target meanings for each word, and the non-target inference probability $a=\frac{3}{4}$. Again we find good agreement, both when the elimination rate $r$ is obtained by directly solving the eigenvalue equation (\ref{eigen}) and by the asymptotic formul\ae\ presented in Appendix~\ref{sec:asymp}. Again the case $\kappa=0$ needs to be treated separately (see Appendix~\ref{sec:kappa0}).

Note that the scale of the decay rate $\kappa$ is different to that considered for the single-word case of Section~\ref{sec:oneword}. As discussed in Section~\ref{sec:moddef}, the mean time between presentations of the same word is $1/\phi_w$ which is equal to $W$ in the case of a uniform word distribution. The memory `half-life' is proportional to $1/\kappa$, so if the lexicon size is increased while keeping $\kappa$ fixed, then the word-meaning associations decay to a much greater extent between presentations of the same word, thereby making it harder to learn. Therefore, to realize lexicon learning times that are accessible in computer simulations, it was necessary to reduce the memory decay rate when exploring multi-word lexicons.

Fig.~\ref{fig:w10m10uniform} also shows learning times when non-target meanings are drawn from a nonuniform distribution distribution. For each word $w$, we rank the $M$ non-target meanings by frequency, so that the $m^{\rm th}$ meaning is inferred with probability
\begin{equation}
\label{invsqrt}
a_{wm} = \frac{a}{\sqrt{m}} \;.
\end{equation}
The same distribution applies for each word. We have chosen this distribution as it belongs to the class where the mean \emph{number} of non-target meanings that is inferred in each episode increases with $M$, the total number of non-target meanings. (Specifically, this number increases as $\sqrt{M}$). Therefore, this choice allows us to make the learning problem arbitrarily difficult, in the sense that the mean number of candidate meanings entertained by the learner at each presentation can be made arbitrarily large.

The prediction from Section~\ref{sec:tlearn} is that the learning time depends only on the largest of the meaning frequencies, which here is $a_{w1}=a$. This prediction relies on the second-largest meaning frequency, $a_{w2}=a/\sqrt{2}$, being sufficiently well separated from the largest meaning frequency that it can be neglected in the calculation of the learning time. We find that by taking $M=1$ in (\ref{tuniform}), with $r$ given by (\ref{r1}), we obtain excellent agreement with the simulation results. This serves as further confirmation that the details of the meaning frequency distribution do not enter into the learning time \cite[a result previously discussed for the case $M\to\infty$ by][]{Blythe2016}.

\begin{figure}
\begin{center}
\includegraphics[width=0.475\linewidth]{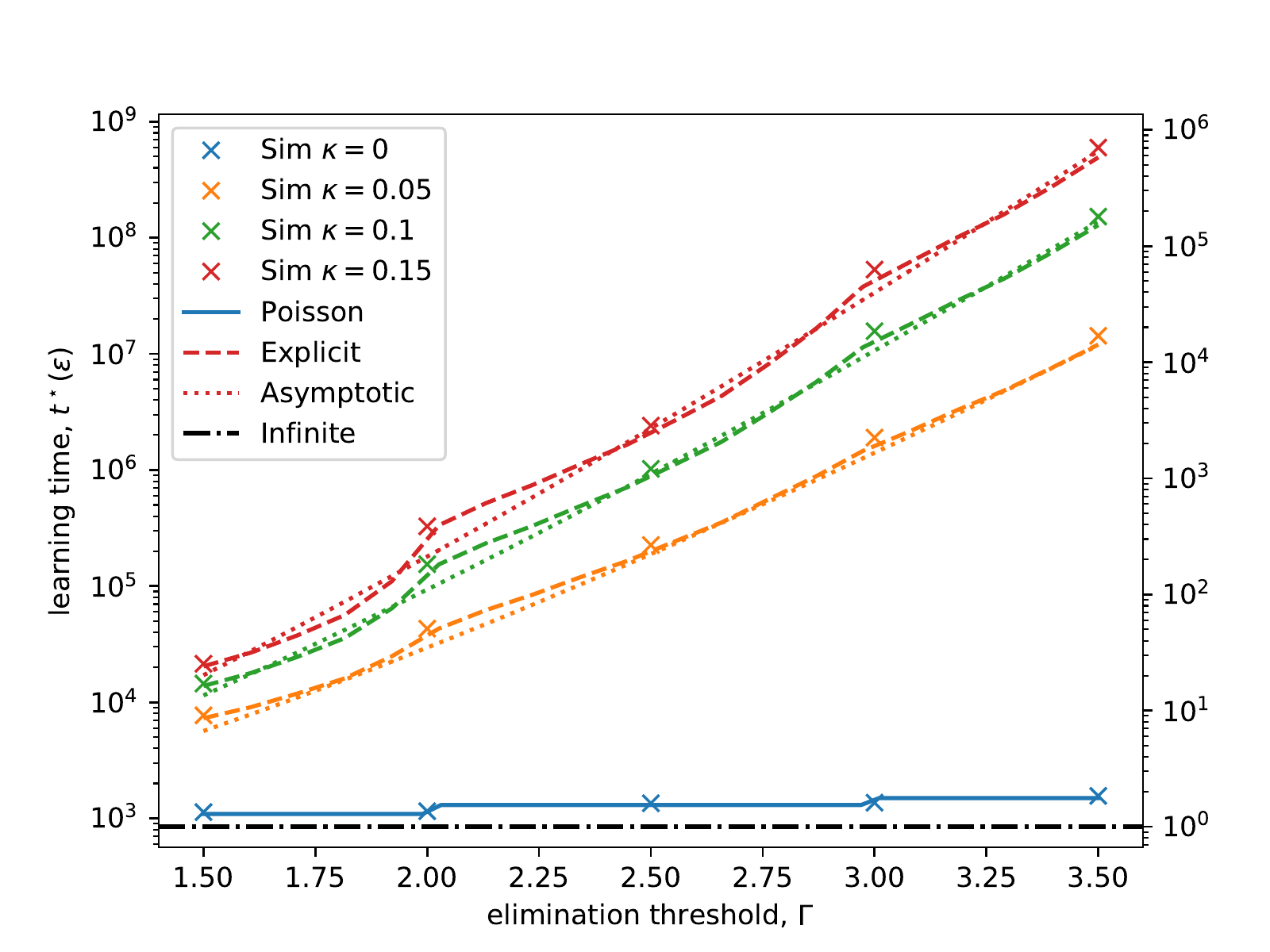}
\includegraphics[width=0.475\linewidth]{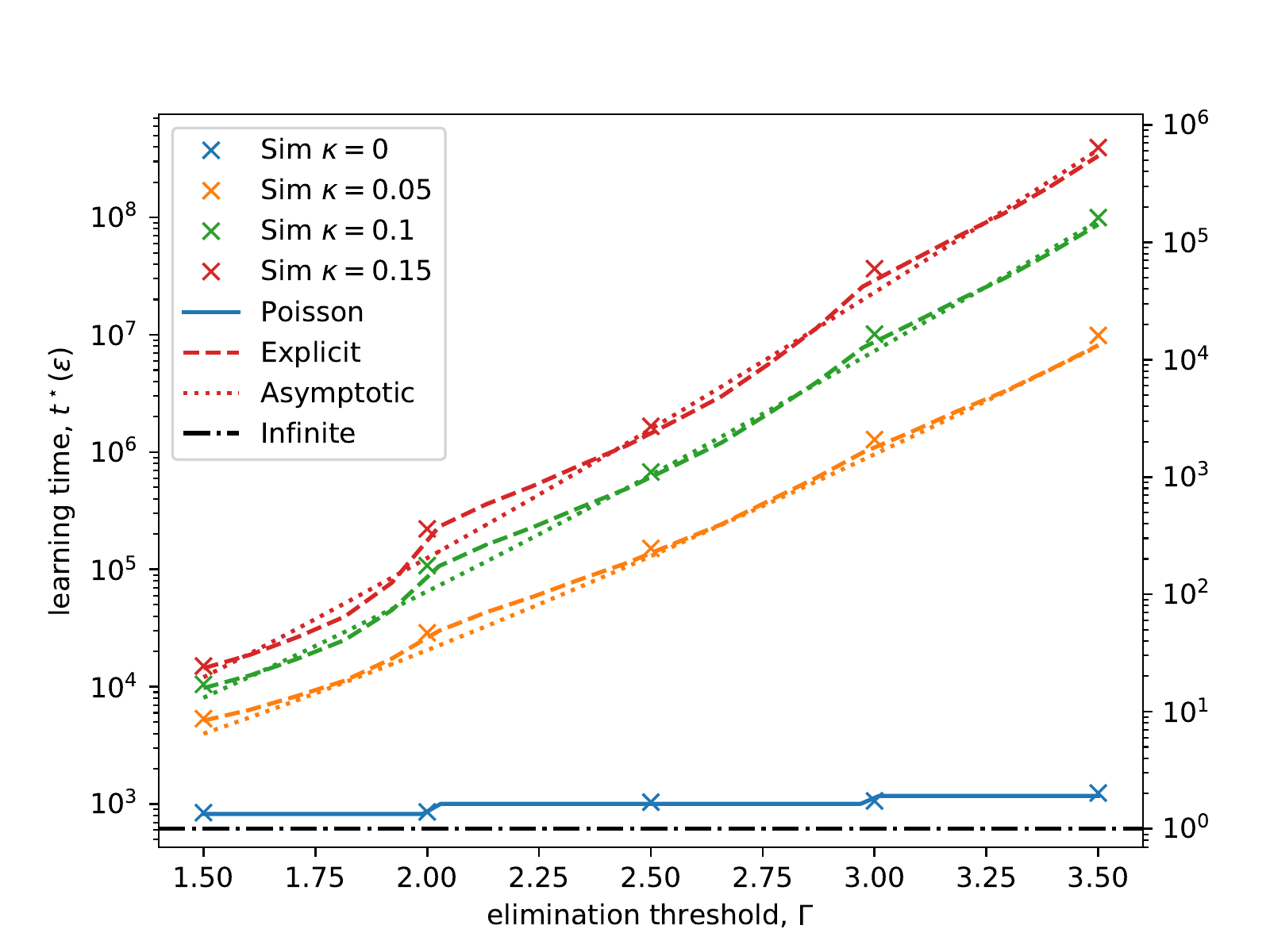}
\end{center}
\caption{\label{fig:w10m10zipf} Same as Fig.~\ref{fig:w10m10uniform} but with a Zipf distribution of word frequencies, Eq.~(\ref{zipf}). Note that due to the greatly extended learning times, a narrower range of elimination thresholds is shown than in the case of a uniform word distribution, Fig.~\ref{fig:w10m10uniform}.}
\end{figure}

In Fig.~\ref{fig:w10m10zipf}, we plot the corresponding learning times when the uniform word distribution is replaced by the Zipf word distribution (\ref{zipf}).  Here, the analytical prediction for the learning time is provided by Eq.~(\ref{tzipf}) rather than (\ref{tuniform}).  The expression involves the rate $r_w$ at which the most frequent non-target meaning is eliminated as a candidate for word $w$, which is obtained from (\ref{r1}) as
\begin{equation}
r_w = \frac{1}{Z} \frac{1}{w} (1 - a) \lambda\left( \frac{Zw \kappa}{1-a} \right) \;.
\end{equation}
Recall that $Z$ here is the normalization of the Zipf distribution, given in Eq.~(\ref{zipf}).  We also require the derivative of this expression with respect to $w$, evaluated at $w=W$. This is given by
\begin{equation}
r_W' = - \frac{1-a}{ZW^2} \left[ \lambda\left( \frac{\kappa}{\phi_W (1-a)} \right) - \frac{\kappa}{\phi_W(1-a)} \lambda'\left( \frac{\kappa}{\phi_W (1-a)} \right) \right] \;,
\end{equation}
which involves the derivative $\lambda'(k)$ of the smallest eigenvalue $\lambda(k)$ with respect to its argument $k$, which can be approximated numerically by a finite difference.

In the case of a uniform distribution over meanings, $a$ is the same for each meaning and $M$ is the number of meanings, in Eq.~(\ref{tzipf}). This provides the predictions for the left-hand plot in Fig.~\ref{fig:w10m10zipf}. For the case of the nonuniform distribution (\ref{invsqrt}) over meanings, we again need keep only the highest meaning frequency, $a$, and put $M=1$. This provides the predictions for the right-hand plot in Fig.~\ref{fig:w10m10zipf}. Again we find excellent agreement between these analytical predictions and simulation results.

\begin{figure}
\begin{center}
\includegraphics[width=0.475\linewidth]{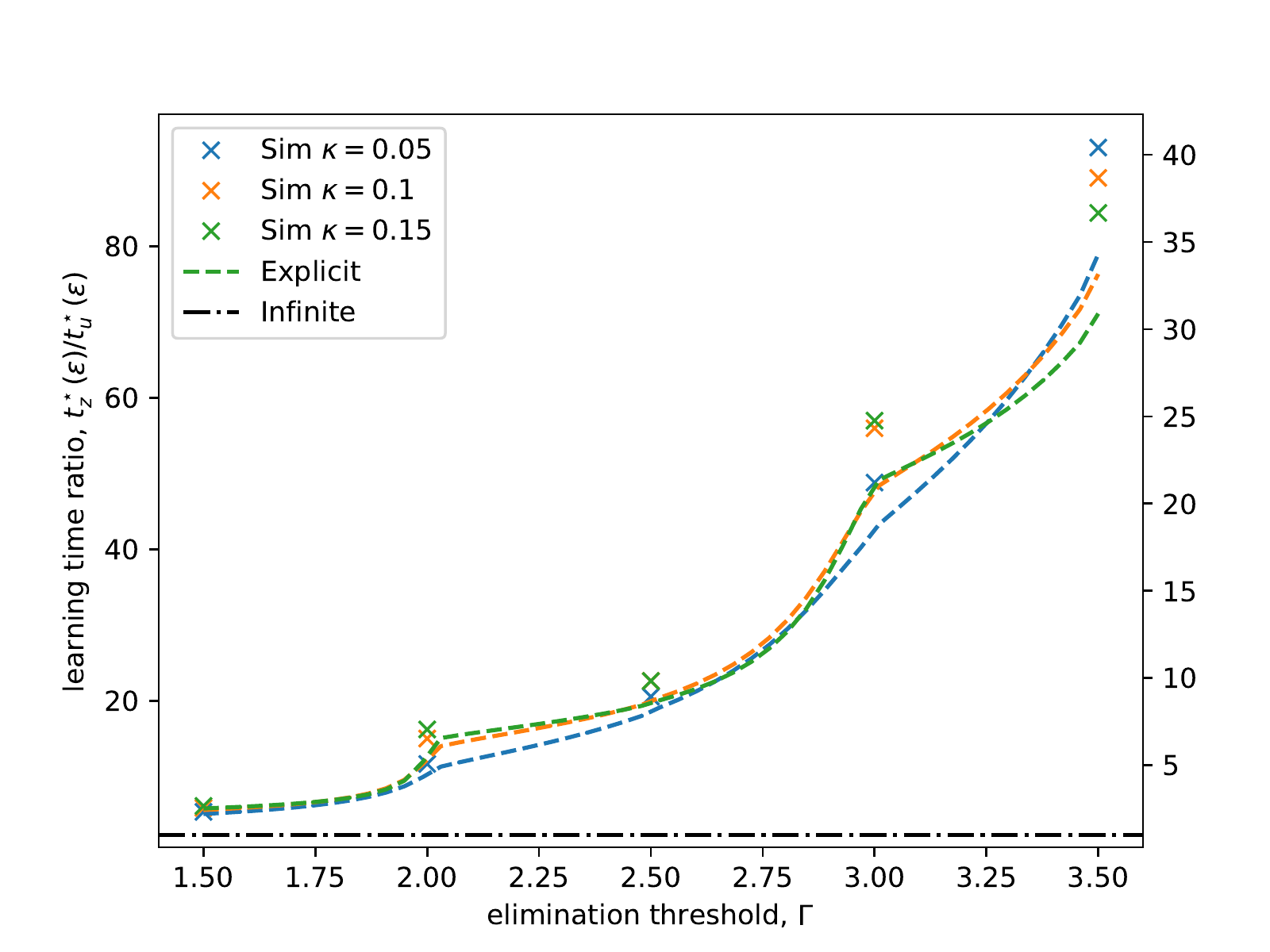}
\includegraphics[width=0.475\linewidth]{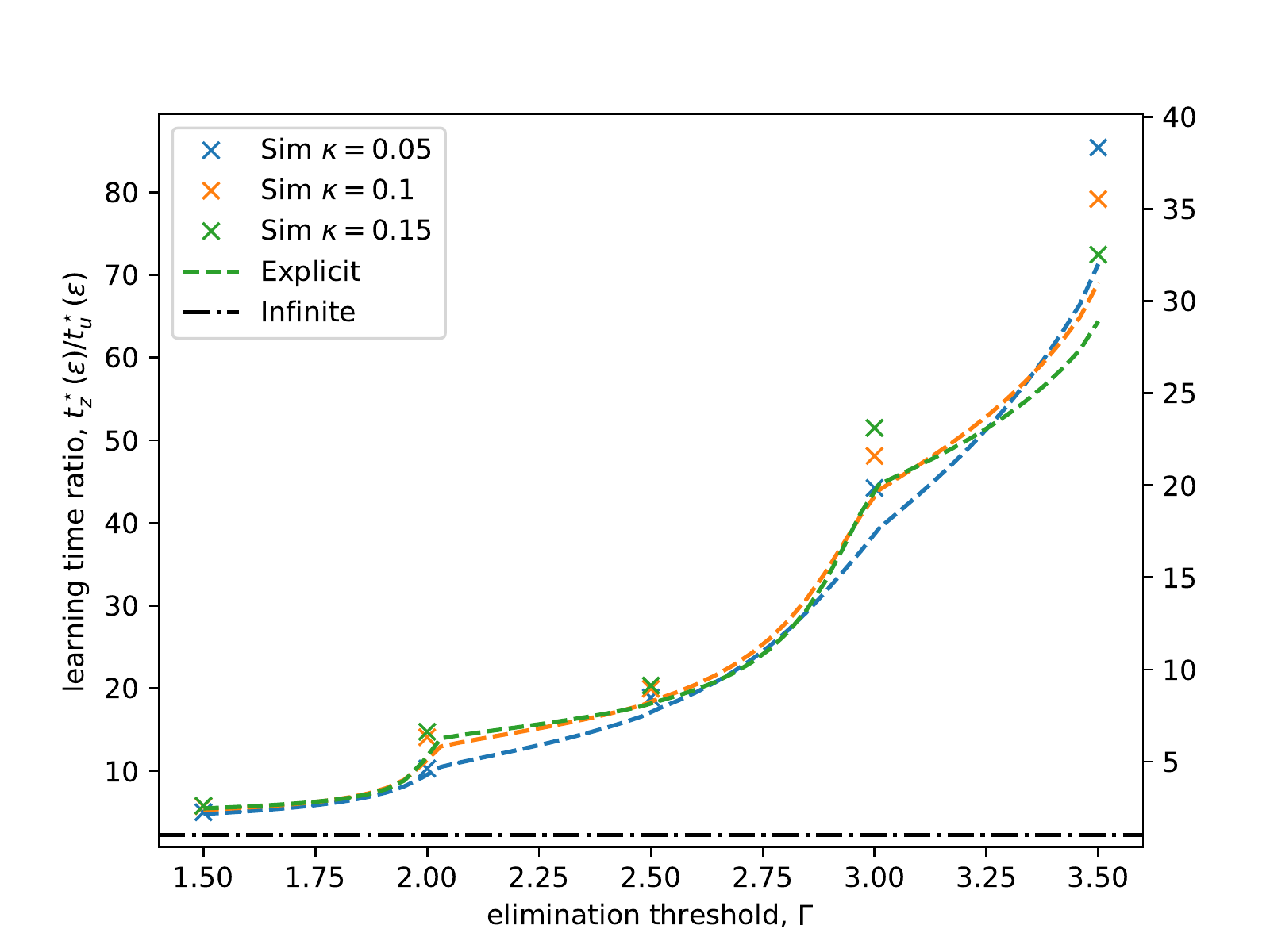}
\end{center}
\caption{\label{fig:w10m10ratio} The ratio of the learning times plotted in Fig.~\ref{fig:w10m10zipf}, where the words have a Zipf distribution, to those plotted in Fig.~\ref{fig:w10m10uniform}, where they have a uniform distribution. The chain line is the corresponding ratio ($Z\approx2.92$) in the case where learners have an infinite memory. Lexicons with a nonuniform word distribution become increasingly hard to learn as the elimination threshold increases.}
\end{figure}

These results illustrate an important difference between learning with infinite and finite memory. With infinite memory, an no elimination threshold (\citealp{Blythe2010}; see also Section~\ref{sec:tlearn}), we found that learning a lexicon with Zipf-distributed words was, in general, a factor of $Z$ longer than learning a lexicon of uniformly-distributed words (assuming the same distribution over meanings). When memory is finite, this result no longer holds. Specifically, with $W=10$ words, we have $Z\approx2.92$, and so one would expect only a modest increase in learning times in Fig.~\ref{fig:w10m10zipf} relative to Fig.~\ref{fig:w10m10uniform}. In fact, we find that learning times increase significantly when words have a Zipf distribution, particularly as the elimination threshold $\Gamma$ increases. This can be seen most clearly from Fig.~\ref{fig:w10m10ratio}, where we plot the ratio of the time to learn a lexicon with a Zipf word distribution to the corresponding time to learn a lexicon with a uniform word distribution. The reason for this slowdown is that rare words are heavily penalized by a finite memory constraint: if the memory lifetime is shorter than the time between successive presentations of the same word, it becomes difficult to eliminate non-target meanings.

A feature that is evident from Figs.~\ref{fig:w10m10uniform}, \ref{fig:w10m10zipf} and \ref{fig:w10m10ratio} are kinks in the learning time at integer values of the threshold $\Gamma$, which are likely to arise from the fact that additional exposures to these rare words are required at these values. Whilst this feature is well captured by the explicit numerical solution of (\ref{eigen}) for the eigenvalue $\lambda(k)$, the asymptotic expression presented in Appendix~\ref{sec:asymp} smooths over this. Nevertheless, the latter continues to give a good approximation. We also see that switching from a uniform meaning distribution to the non-uniform distribution (\ref{invsqrt}) in general has a more limited effect on the learning time than changes in the word distribution or other parameters, as the analysis of Section~\ref{sec:tlearn} predicts.

\subsection{Mutual exclusivity constraint}
\label{sec:mutex}

In this Section we show that knowledge of the late-time rate $r_{wm}$ at which the non-target meaning $m$ is excluded as a candidate meaning for word $w$ also allows us to determine the time required to learn a lexicon when a mutual exclusivity constraint \citep{Markman1988} is operating. The specific form of this constraint is as described by \cite{Reisenauer2013}: as soon as one word is learnt, its meaning is immediately eliminated as a candidate meaning for all other words. One way to think of this is that the threshold criterion for eliminating a non-target meaning is immediately met, by virtue of that meaning having been assigned to some other word. Note that this elimination by mutual exclusivity can cause other words to be learnt (if there were no other remaining non-target candidate meanings at that time), which in turn may have a knock-on effect and allow further words to be learnt. Indeed, this constraint in principle allows words to be learnt on their first exposure (i.e., fast mapped). A simulation algorithm that implements this constraint is laid out in Appendix~\ref{sec:algos}.

\cite{Reisenauer2013} showed that two factors contribute to the probability $L(t)$ that a lexicon has been acquired by time $t$. The first of these factors is the probability that the learner has encountered every word in the lexicon at least once, which is clearly a necessary requirement to learn the lexicon. It also corresponds to the case where all words are fast mapped. This factor is given by
\begin{equation}
\label{LFM}
L_{\rm FM}(t) = \prod_{w=1}^{W} \left[ 1 - {\rm e}^{-\phi_w t} \right] \;.
\end{equation}

The second factor is more subtle. As candidate meanings are eliminated by crossing the relevant threshold, the situation may arise that only one unique assignment of meanings to words is compatible with the mutual exclusivity constraint (i.e., that no two words can have the same meaning). The simplest case where multiple assignments of meanings to words is possible is if the target meaning of word $w$ (that is, $\target{w}$) is a candidate for some other word $w'$, and the meaning of $w'$ ($\target{w'}$) is a candidate for $w$. In this case, we cannot decide if word $w$ has meaning $\target{w}$ or $\target{w'}$, and likewise for word $w'$. The rate at which the ambiguity between such a pair is eliminated is $r_{w\target{w}} + r_{w'\target{w'}}$, since establishing the meaning of $w$ may allow that of $w'$ to be established, or vice versa.  There may be larger groups of mutually ambiguous words (e.g., triplets); however ambiguity between them is eliminated more rapidly than between pairs of words. Consequently, the second factor that contributes to the probability that the lexicon is learnt by time $t$ is
\begin{equation}
\label{Lpairs}
L_{\rm pairs}(t) = \prod_{\langle w, w'\rangle} \left[ 1 - {\rm e}^{-(r_{w \,\target{w'}} + r_{w' \,\target{w}})t} \right]
\end{equation}
where the product is over all distinct pairs of words $w,w'$.  The total probability that the lexicon has been learnt under mutual exclusivity is then
\begin{equation}
\label{Ltme}
L_{\rm ME}(t) \approx L_{\rm FM}(t) L_{\rm pairs}(t) \;.
\end{equation}

We obtain a learning time in the same way as before, that is, by keeping only the slowest decaying pairs in (\ref{Lpairs}) and then solving $L_{\rm ME}(t) = 1 - \epsilon$.  We explain the details of this procedure in Appendix~\ref{sec:meuni}. Here we focus on the situation where the mutual exclusivity constraint has the biggest impact on learning times, which is when the frequency of a word is positively correlated with the frequency that its target meaning is inferred as a candidate for other words. The reason why the mutual exclusivity constraint has the greatest power under these conditions is that it is high-frequency non-target meanings that limit the rate a word can be learnt. These high-frequency confounding meanings are eliminated early on in the learning process if they correspond to high-frequency words, since high-frequency words tend to be encountered earlier than low-frequency words (and hence learnt more rapidly).

Specifically, we assume that the ranks of the words and their target meanings are the same (i.e., $\target{w}=m$), that the word with rank $w$ has a frequency given by the Zipf distribution (\ref{zipf}) and that the inference frequency $a_{wm}$ has the slightly modified inverse square root form
\begin{equation}
\label{invsqrtme}
a_{wm} = \left\{ \begin{array}{ll}
\frac{a}{\sqrt{m}} & 1 \le m < w \\
\frac{a}{\sqrt{m-1}} & w < m \le W
\end{array} \right. \;.
\end{equation}
This expression takes into account that the target meaning ($w=m$) should not be included in the ranking of non-target meanings, and in particular that the most frequent non-target meaning always has frequency $a$, as has been our convention throughout this work. Note that we also assume that target meaning of any word can be inferred in the context of any other word, with the frequency specified above, and that no other meanings are entertained as candidates. 

In the case of learning with infinite memory \citep{Reisenauer2013}, we found that the slowest decaying contribution to (\ref{Ltme}) came either from waiting for all words to be exposed at least once, i.e., the factor $L_{\rm FM}(t)$, or from resolving ambiguity between the two most frequent words, i.e., the term $w=1, w'=2$ in (\ref{Lpairs}). That is, in the latter instance, one would need to wait until either $m=2$ has been excluded as a candidate for $w=1$, or $m=1$ as a candidate for $w=2$, by cross-situational learning alone. Once this had been done, these frequent non-target meanings would then be eliminated as candidates for all other words. Since all other non-target meanings have a much lower frequency, they are very likely to have been eliminated through non-appearance in the time it has taken to resolve the most confusable pair of words.

The situation changes when learners have a finite memory. The difference is that low-frequency non-target meanings of low-frequency words become harder to eliminate, as the evidence that is needed to do so is more likely to be forgotten between exposures of these rare words. Now, it is more efficient to eliminate these low-frequency non-target meanings by application of the mutual exclusivity constraint, and the learning time is now determined by the resolution of ambiguity between pairs of words in the low-frequency parts of the word and meaning spectrum. Necessarily, this takes longer than the time required to expose each word individually; therefore the factor $L_{\rm FM}(t)$ in (\ref{Ltme}) can be neglected.   Since there are many pairs of words in the low-frequency part of the spectrum that have a similar elimination rate ($r_{ww'} + r_{w'w}$) we cannot simply keep the single slowest decay in (\ref{Lpairs}). The mathematical procedure for doing this is set out in Appendix~\ref{sec:meuni}. The resulting prediction for the learning time is
\begin{equation}
\label{tastme}
t^\star = \frac{1}{\tilde{r}} \Lambert \left( \frac{\tilde{r}}{|\tilde{r}'|} \frac{1}{\sqrt{2\epsilon}} \right) \;.
\end{equation}
where $\Lambert(z)$ is again the Lambert W function that appeared in Eq.~(\ref{tzipf}).  Here, the quantities $\tilde{r}$ and $\tilde{r}'$ are given by
\begin{align}
\tilde{r} &= r_{W,W-1} = \frac{1}{WZ} \left(1-\frac{a}{\sqrt{W}}\right) \lambda(k) \\
\tilde{r}' &= \left. \frac{{\rm d}}{{\rm d} w} (r_{wm}+r_{mw}) \right|_{w=W, m=W-1}
= -\frac{1}{W^2Z} \left[ 1-\frac{3a}{2\sqrt{W}} \right] \left[ \lambda(k) - k \lambda'( k ) \right] \;.
\label{endtastme}
\end{align}

\begin{figure}
\begin{center}
\includegraphics[width=0.45\linewidth]{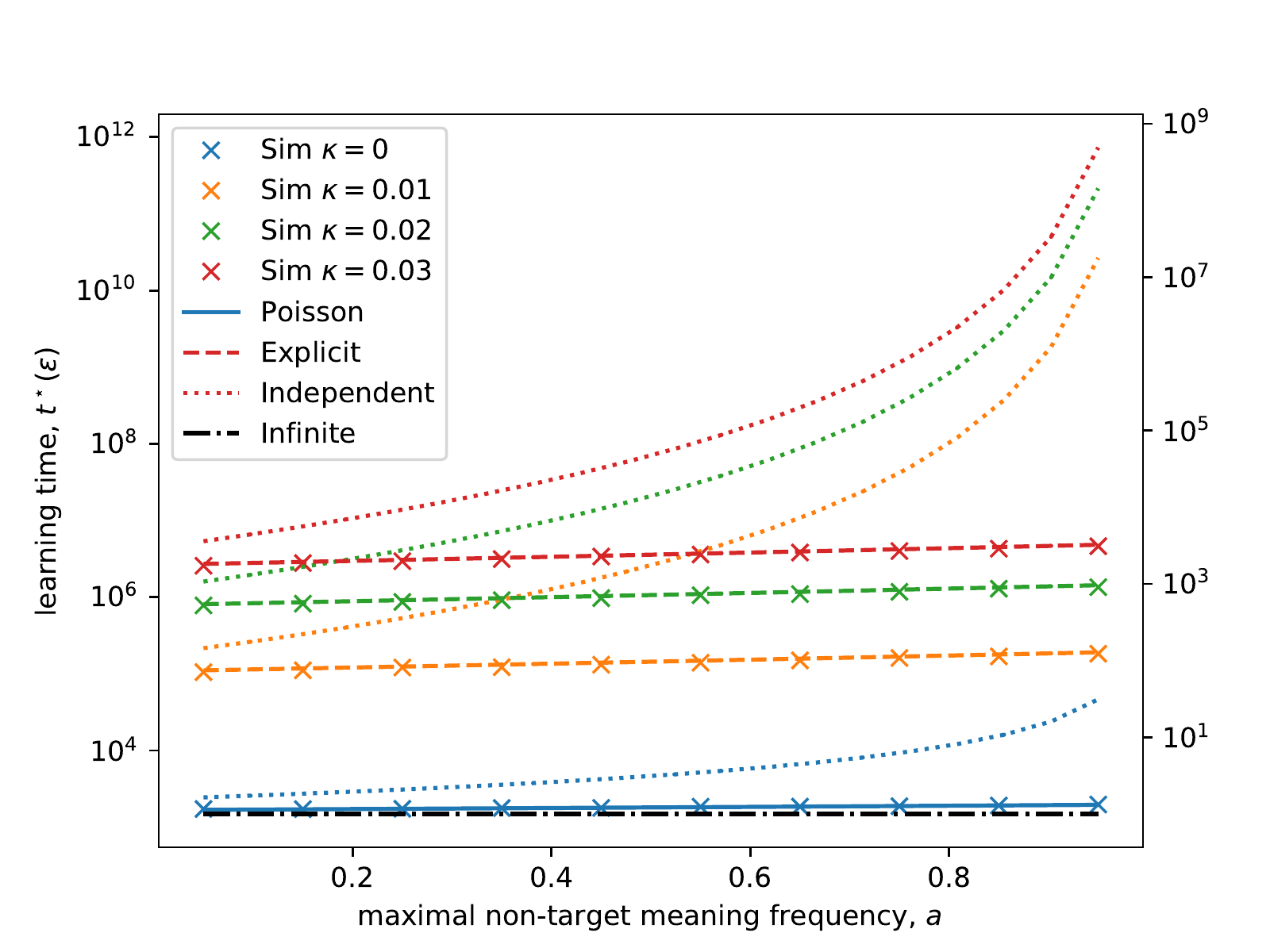}
\includegraphics[width=0.45\linewidth]{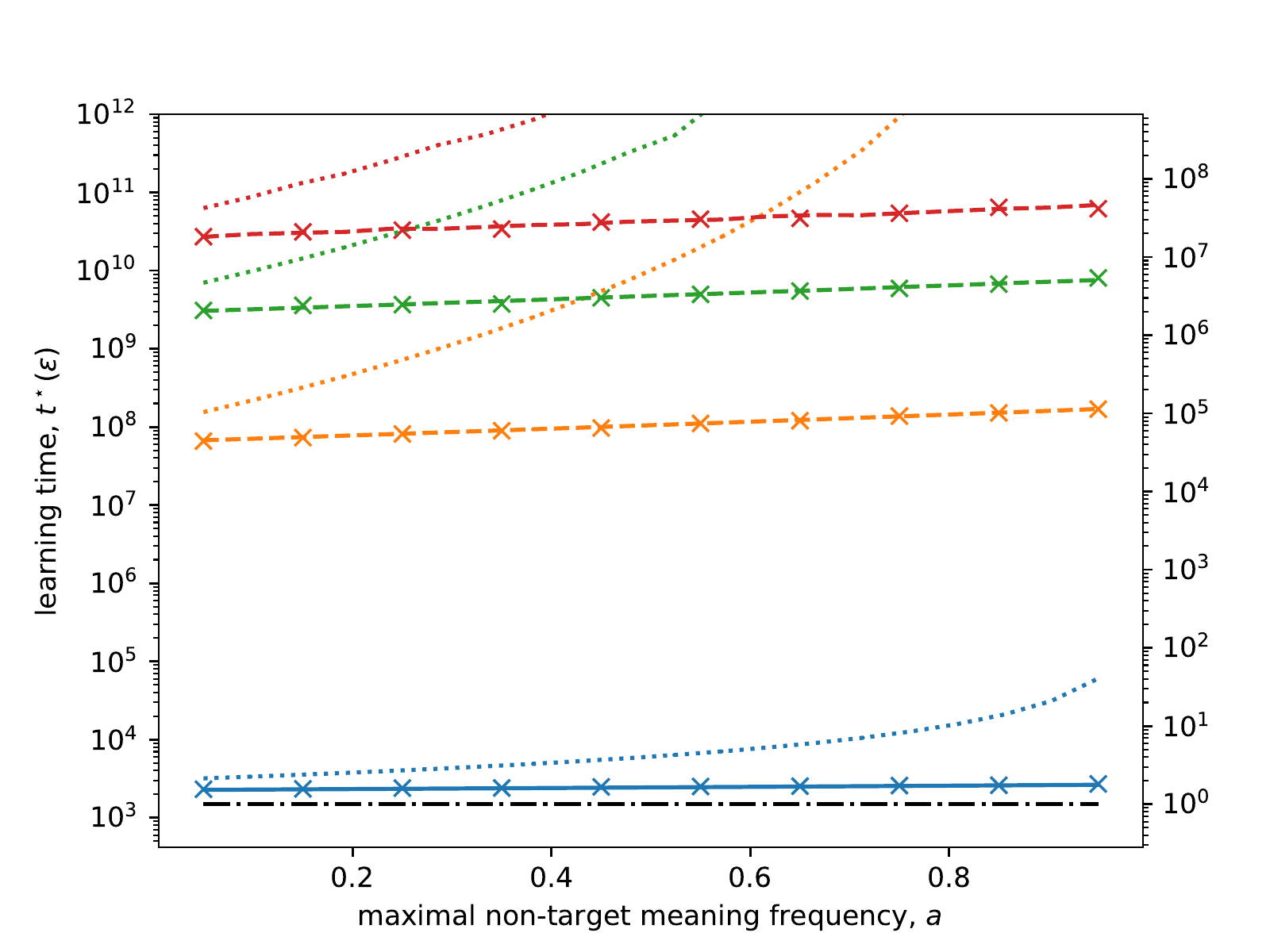}
\end{center}
\caption{\label{fig:w50exclude} Learning time for a lexicon of $W=50$ words with a mutual exclusivity constraint. The words have a Zipf frequency distribution (\ref{zipf}) and the non-target meanings are distributed according to (\ref{invsqrtme}) where the amplitude $a=\frac{3}{4}$ and the exponent $\gamma=\frac{1}{2}$. Left panel: elimination threshold $\Gamma=3$; Right panel: elimination threshold $\Gamma=5$. Crosses show points from between 100 and 10,000 Monte Carlo simulations of the process with different memory decay rates $\kappa$ (longer times imply smaller samples). Solid line is the learning time prediction using the Poisson distribution for $\kappa=0$; dashed lines are obtained from (\ref{tastme}) via numerical solution of the eigenvalue equation (\ref{eigen}). Dotted lines indicate the learning times without mutual exclusivity operation; the chain indicates the case with $\Gamma<1$.}
\end{figure}

We plot the prediction (\ref{tastme}) for the learning time in Fig.~\ref{fig:w50exclude} for the case of a lexicon of $W=50$ words, and again find good agreement with those obtained from Monte Carlo simulations of the process. Whilst lexicon learning is somewhat slower than when learners have infinite memory---particularly for a large elimination threshold $\Gamma$---we find that mutual exclusivity still significantly accelerates learning. This is particularly the case when referential uncertainty is high: in fact, when mutual exclusivity is operating, the learning time is insensitive to referential uncertainty, except when the most frequent non-target meaning has an appearance frequency very close to unity \citep{Reisenauer2013}. One consequence of this accelerated learning is that the asymptotic expressions (\ref{lasymp}) do not give a good approximation to the learning time under mutual exclusivity; hence these predictions are omitted from Fig.~\ref{fig:w50exclude}.

\section{Human-scale lexicons}
\label{sec:60k}

In the previous two sections we obtained mathematical predictions for the lexicon learning time that are based on the smallest eigenvalue of equation (\ref{eigen}), and validated these against Monte Carlo simulations for small lexicons. The utility of these expressions is that we can now apply them with confidence to lexicons and sequences of exposures of the scale encountered by humans, as simulations at this scale become prohibitive.  As in earlier work \citep{Blythe2010}, and following \cite{Bloom2000}, we take the representative size of the lexicon to be $W=60,000$ words.  The other key figure is the number of words that a child is exposed to over an $18$-year period. Estimates of the word exposure rate vary from around $600$ to $3,000$ words per hour, depending on social class and whether one includes only child-directed speech, or also takes ambient speech into account \citep{Hart1995,Sperry2018}. There is also the question of the how many hours a day a child will be attending to these word productions. We consider therefore two scenarios: (A) a mid-point word rate of $1,800$ words per hour, combined with $12$ hours of attention per day, which corresponds to total exposure of $1.42\times 10^8$ words during the acquisition period; and (B) a more conservative estimate of $600$ word per hour and an $8$-hour day, which lies at the lower end of the range reported in the literature, and corresponds to $3.16\times10^7$ total exposures.

It is reasonably well established that word forms have a Zipf distribution (\ref{zipf}), at least up to the lexicon size considered here \citep{Ferrer_i_Cancho2001}, and so we adopt this distribution in this investigation. However, we have little empirical guidance as to what appropriate choices for the meaning frequency distribution $a_{wm}$, the elimination threshold $\Gamma$ or memory decay rate $\kappa$ should be. Therefore our strategy is to leave these as free parameters, and determine combinations of them that allow the lexicon to be learnt on the requisite timescale. Matters are simplified somewhat if we assume a nonuniform meaning distribution (which seems reasonable), as we have confirmed that only the most frequent non-target meaning is relevant to the learning time when words are learnt independently.  For simplicity, we take this maximal frequency to have the same value, $a$, across all words.

\begin{figure}
\begin{center}
\includegraphics[width=0.475\linewidth]{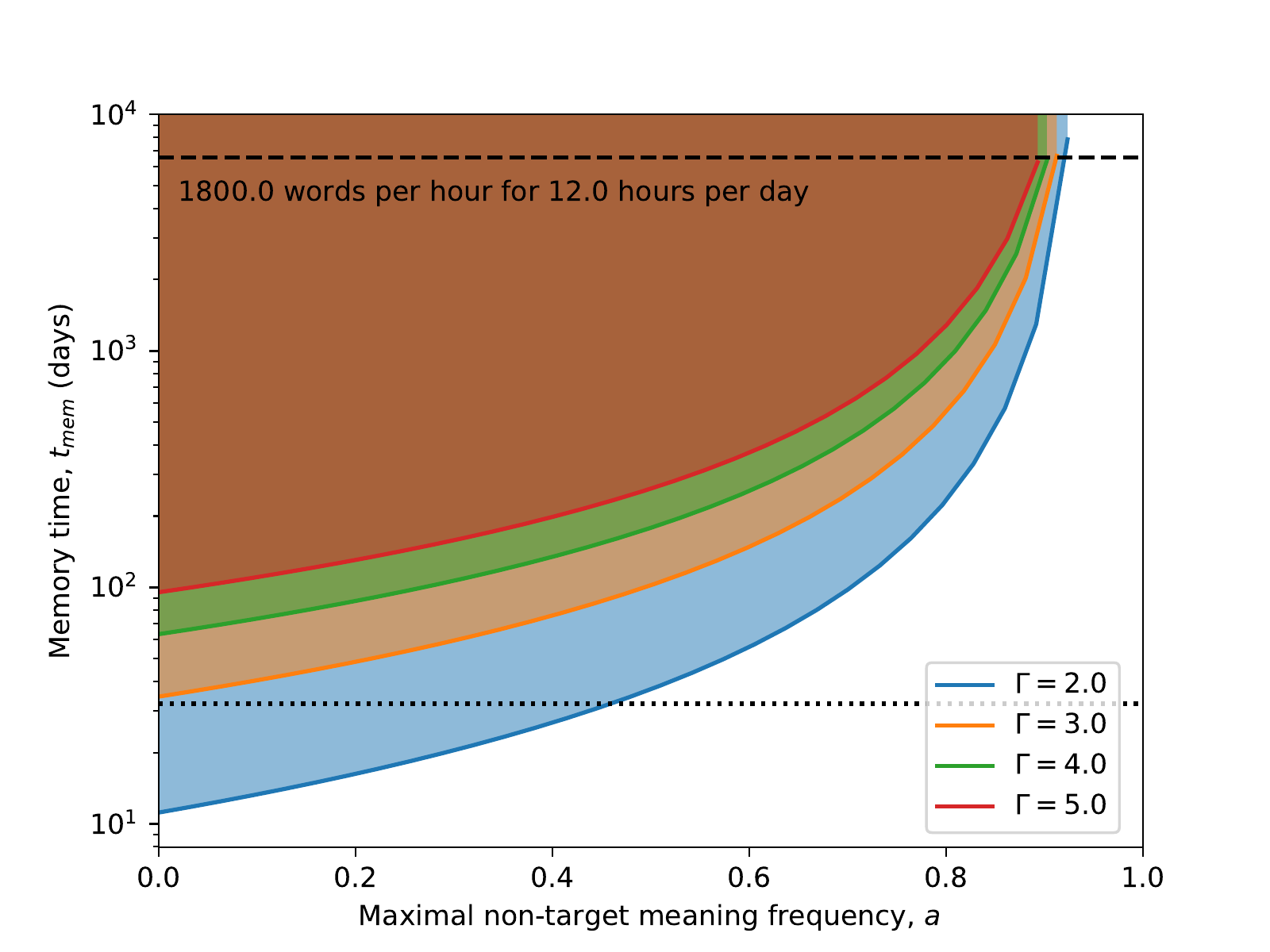}
\includegraphics[width=0.475\linewidth]{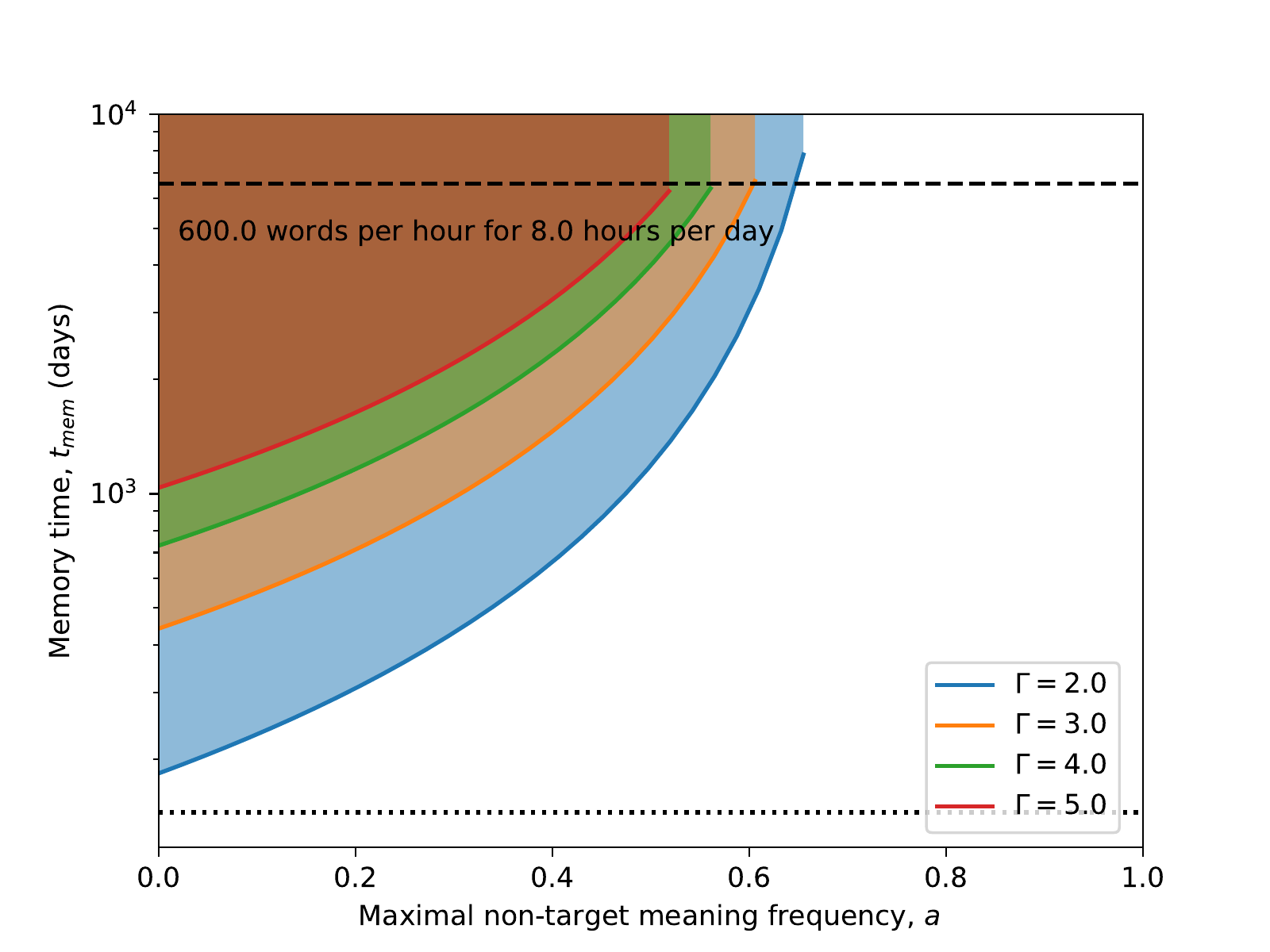}
\end{center}
\caption{\label{fig:pd-ind} Parameter space over which a lexicon of $W=60,000$ Zipf-distributed words can be learnt independently within two scenarios of total childhood word exposures. Left figure (scenario A): $1,800$ words per hour, for $12$ hours a day over $18$ years. Right figure (scenario B): $600$ words per hour, for $8$ hours a day over $18$ years. Each shaded region shows the range of memory lifetimes ($1/\kappa$ in days) and maximal non-target frequencies $a$ over which the lexicon can be learnt by all but a fraction $\epsilon=0.01$ of the population for a given setting of the elimination threshold $\Gamma$. These regions become smaller as $\Gamma$ is increased. The lower dotted line indicates the mean times between presentations of the least common word (scenario A: 32 days; scenario B: 145 days). The upper dashed line indicates a mean memory lifetime that is equal to that of the $18$-year learning period.}
\end{figure}

If Fig.~\ref{fig:pd-ind}, we plot the range of memory lifetimes ($1/\kappa$, in days) over which the lexicon can be learnt as a function of the frequency $a$ and for various values of elimination threshold $\Gamma$ under the two word-exposure scenarios given above, all under the assumption that all words are learnt independently. The shaded region indicates which combinations of memory lifetime and non-target meaning frequency are viable for different settings of the elimination threshold $\Gamma$.  The boundary of each region is determined numerically by varying $\kappa$ until the predicted value of $t^\star$ equals the lifetime exposure value. Note that there is an upper limit on the amount of referential uncertainty (quantified by the probability $a$) where the lexicon can be learnt in the requisite time, even with infinite memory.  In calculating $t^\star$ we employ explicit numerical solution of the equation (\ref{eigen}), as this turns out to give more precise results overall than the asymptotic expression (\ref{lasymp}) with only a modest computational effort. We note that the assumption that only the smallest eigenvalue of (\ref{eigen}) contributes to the overall learning time may start to break down when the memory lifetime becomes of order of the 18-year acquisition period, and precise location of the boundary may be unreliable in this region. However, we are most interested in the regime where memory lifetimes are somewhat shorter than this, and in this regime errors arising from this approximation can be neglected.

To interpret this plot, we need to determine a reasonable limits on a human memory lifetime for word learning. Artificial word-learning experiments suggest that $18$-month olds are capable of retaining the meaning of a learnt word for around $10$ weeks \citep{Wojcik2013}. This suggests a value of around $100$ days as an order-of-magnitude limit on a reasonable memory lifetime: whilst older children may have longer retention periods, it is also true that this research pertains to retention of learnt words, rather than associations between words and potential meanings of unlearnt words, which may decay more rapidly.  Under the mid-point word-exposure scenario A, we find that the lexicon can be acquired within this limit as long as the degree of referential uncertainty (i.e., the maximum probability $a$ of inferring a non-target meaning) is less than about $0.8$, when the elimination threshold $\Gamma$ is low, to near zero once $\Gamma$ reaches a value of $5$. This is to be compared with a referential uncertainty of about $0.9$ that can be tolerated when memory is infinite. It is however interesting to note that the lexicon is still learnable when the memory lifetime is less than mean time between exposures of the least frequent word (shown by the dotted line in the figure): this is presumably due to fluctuations that generate short bursts of the same word within the memory time window. Under scenario B where the rate of word exposure is somewhat lower, the constraints on the learnability of the lexicon are more severe, and would require memory lifetimes of a couple of years in order to tolerate even low degrees of referential uncertainty.  Since the lowest word rate arises by considering only child-directed speech from a primary caregiver \citep{Hart1995}, we find that the capacity for a large lexicon to be learnt by cross-situational learning alone depends on the extent to which other sources of linguistic data can be exploited by children.

\begin{figure}
\begin{center}
\includegraphics[width=0.475\linewidth]{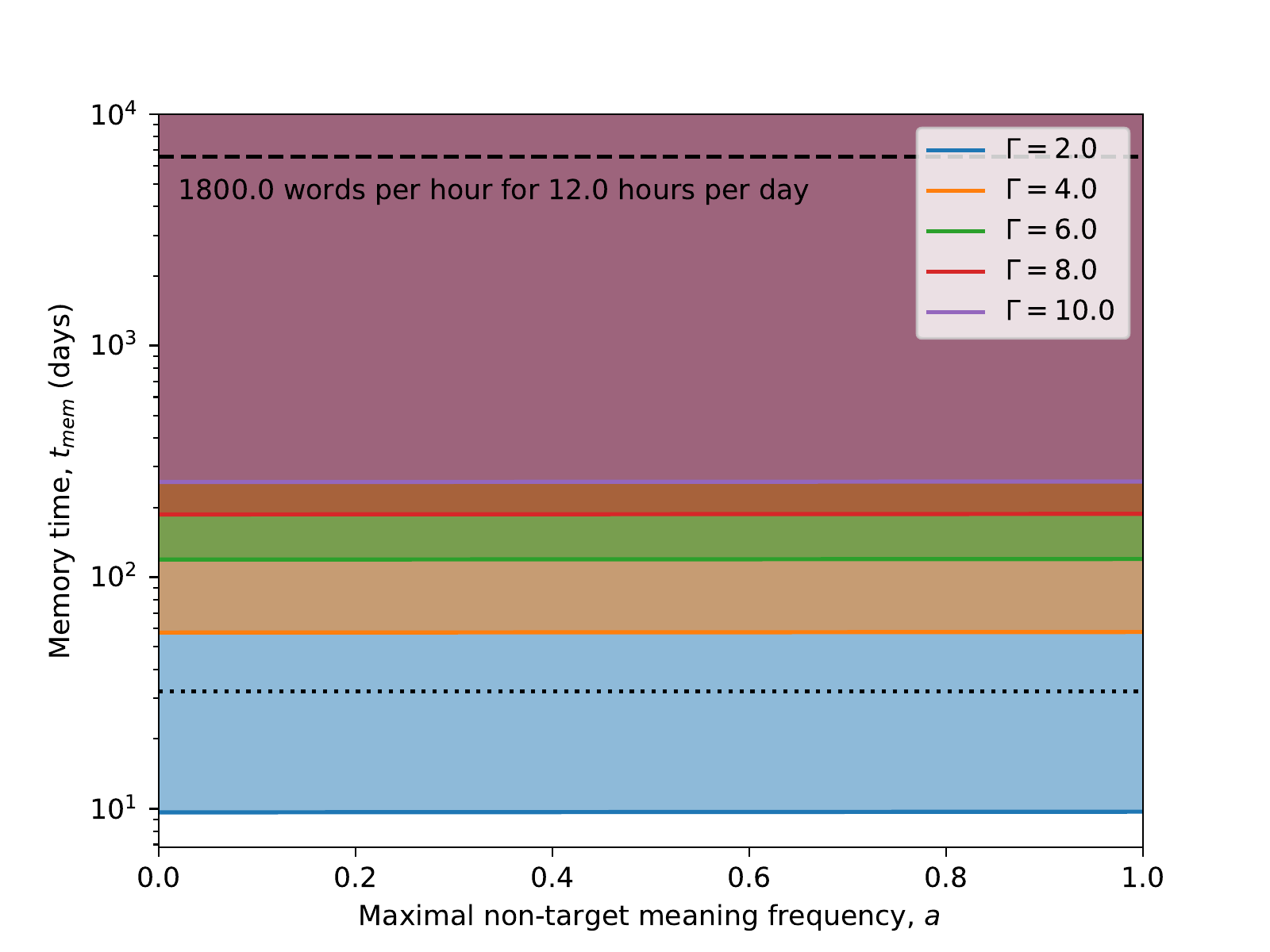}
\includegraphics[width=0.475\linewidth]{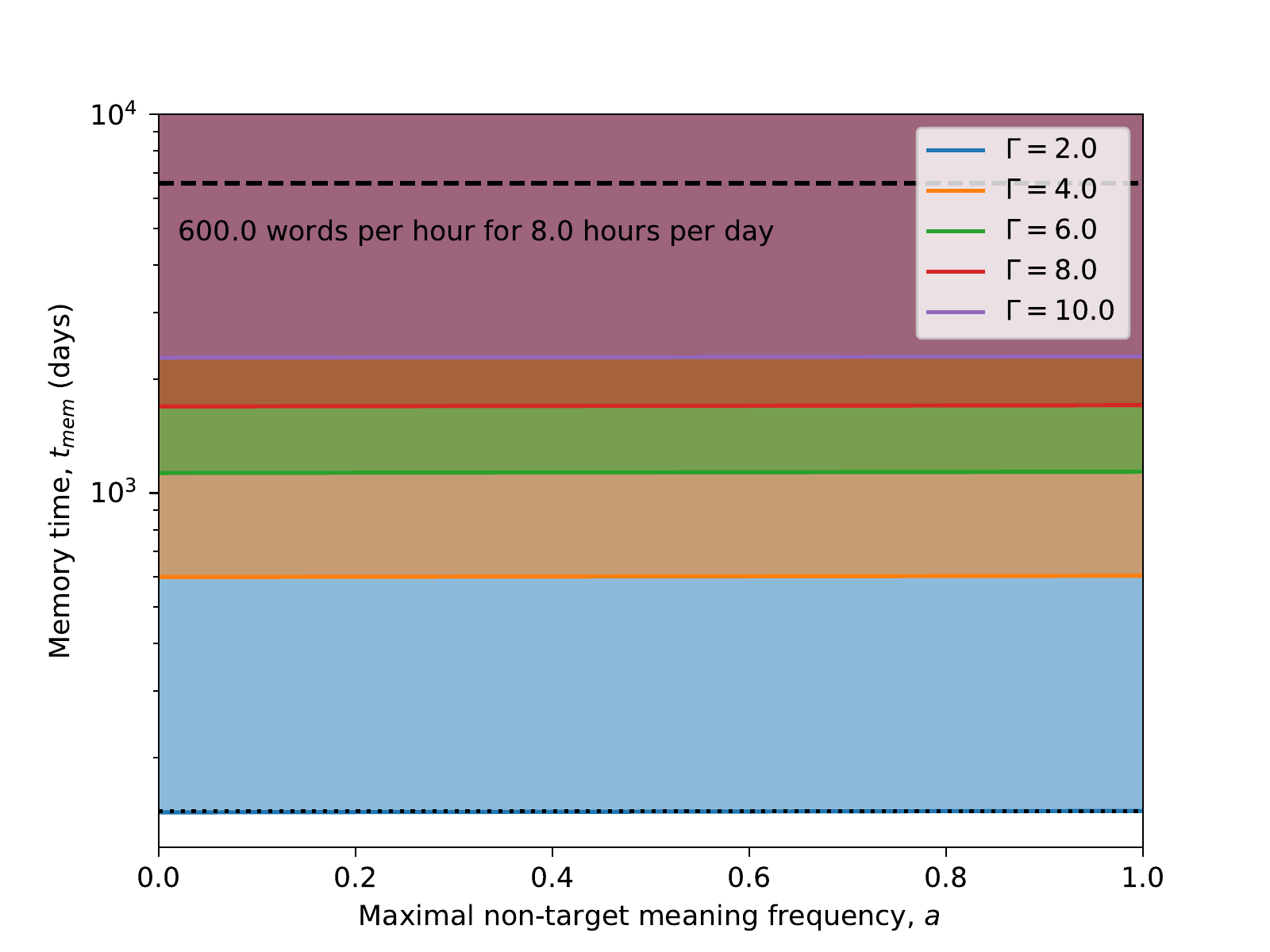}
\end{center}
\caption{\label{fig:pd-me} Parameter space over which a lexicon of $W=60,000$ words can be learnt under the application of a mutual exclusivity constraint within the same two scenarios as in Figure~\ref{fig:pd-ind}. The meaning distribution is of the inverse square-root form (\ref{invsqrtme}). The mutual exclusivity constraint almost entirely eliminates the dependence on the degree of referential uncertainty.}
\end{figure}

In Fig.~\ref{fig:pd-me} we investigate the version of the model with the mutual exclusivity constraint operating. Here, we do need to specify a form for the meaning distribution, due to the learning time being determined by resolving ambiguity between the least frequent words. We adopted the same inverse square root law (\ref{invsqrtme}) as previously, and find that the learning time is insensitive to the amplitude $a$ of this distribution. We have also found that other meaning frequency distributions (not shown) give essentially the same result, which means that our uncertainty as to what this distribution should be does not strongly affect our assessment of the capability of the model learner to acquire a human scale lexicon under mutual exclusivity. Specifically, we find that the main restriction on the model parameters for this to be achievable is that the the elimination threshold is not too large, that is $\Gamma<5$ in the case of Scenario A, and $\Gamma<2$ under scenario B.  This shows that a pure cross-situational learner, endowed with an additional mutual-exclusivity constraint, can acquire realistically large lexicons under a fairly wide range of model assumptions.

\section{Discussion}
\label{sec:disco}

In this work, our aim was to understand how a simple model of cross-situational word learning, which is based on tracking the co-occurrence of words and possible referents, performs when the assumption of infinite memory capacity that was made in earlier studies \citep{Smith2006,Blythe2010,Reisenauer2013,Blythe2016} is relaxed. There are two main new features in this model. First, the strength of association between a word and a candidate meaning decays exponentially at rate $\kappa$. Second, the difference in two association strengths must exceed a threshold $\Gamma$ in order for a candidate meaning to be eliminated. Thus, in this model, a single occasion where a word is presented and a particular meaning is \emph{not} inferred as a potential referent is insufficient for the learner to discard the meaning as a candidate, as it was in earlier models. Both infinite memory and sensitivity to inferences drawn in a single instance of use are perhaps unreasonable expectations of real word learners \citep[see e.g.][for a discussion of related concerns]{Medina2011}.

Our principal finding is that although both modifications to the model lead to longer word-learning times (as expected), it is still possible for this model learner to acquire a lexicon of $60,000$ words \citep{Bloom2000} after a number of exposures consistent with some estimates of the number of words encountered in an $18$ year period \citep{Hart1995,Sperry2018}. However, there are limits on the degree of referential uncertainty that can be tolerated by a learner, and on the amount of evidence that is required to exclude a candidate meaning from consideration, and both become somewhat severe if the number of word exposures available to the model learner lies at the lower end of the range reported in the literature.

More precisely, if we consider an amount of exposure in the middle of the range of reported values \citep{Hart1995,Sperry2018}, then with a fairly low elimination threshold $\Gamma=2$, a learner needs to exclude (by word learning heuristics, for example) alternative meanings only around $20\%$ of the time to acquire a $60,000$-word lexicon by adulthood. Throughout this work, we have seen that learning times are particularly sensitive to the threshold $\Gamma$: beyond a value of about $5$, the model learner cannot acquire the lexicon under reasonable constraints on its memory lifetime. However, we have assumed that words are presented as independent Poisson processes, and in reality it may be that particular words come in bursts \citep{Altmann2009} which would allow the elimination threshold to be crossed more easily.  The capacity to learn a lexicon is also strongly dependent on the sheer number of word exposures that are available to a child for the purpose of identifying a meaning. We found that a model learner exposed to words at the lowest rate reported for child-directed speech from a primary caregiver struggled to acquire a lexicon of $60,000$ words.  This raises important questions about what aspects of linguistic behavior in a child's environment provide opportunities for word-meaning associations to be established.  Moreover, it is unclear whether it is reasonable to apply a target lexicon size of $60,000$ words universally across all learners. If we ask instead for a lexicon of only half this size to be acquired by adulthood, then we find that this can be achieved in the low word-exposure scenario with levels of referential uncertainty and elimination thresholds that could be tolerated when acquiring the larger lexicon at the faster word-exposure rate. It is also possible that memories of different word-meaning associations decay at different rates: for example, it might be the case that episodes of use where only a small number of candidate meanings are inferred decay less rapidly than when many meanings are competing.

In any case, as we found in the context of the infinite-memory learner \citep{Reisenauer2013}, a mutual exclusivity constraint \citep{Markman1988} provides a powerful mechanism for acquiring a lexicon. In fact, this constraint allows a learner to acquire a large lexicon with almost any distribution of candidate meanings, at least, under the assumption that the frequencies of non-target meanings are positively correlated with the frequencies of associated words. Under these conditions, the most likely alternative meanings for a low-frequency word will be out of contention when such words are encountered for the first time, which in turn may allow them to be fast mapped (i.e., acquired on a first exposure).

Crucially, all the different conditions described above share the property that a meaning is only assigned to a word (i.e., learnt) once all other possible meanings have been excluded, either through the accumulation of sufficient negative evidence, or the mutual exclusivity constraint. This is to be contrasted with accounts in which a specific meaning is hypothesized at the first encounter with a word, and the hypothesis modified over time if it is found to conflict with the evidence presented to the learner \cite[see e.g.][]{Smith2011,Trueswell2013}.  It has been suggested that difficulties encountered by real-world word learners, such as many different words appearing between two presentations of the same word, or finite-memory constraints, favor such hypothesis-testing accounts \citep{Medina2011}. The model discussed here explicitly incorporates some of these complications, and although they can have a significant effect on the time to acquire a lexicon, it may not be to such a great extent that we are forced to appeal to additional word-learning mechanisms, like hypothesis-testing.

Nevertheless, it may be that real-world learners \emph{do} combine cross-situational statistics and hypothesis testing, something that was considered for a model learner with infinite memory by \cite{Blythe2010}.  Translated to the model discussed here, this specific hypothesis-testing strategy would amount to a learner randomly choosing from the set of candidate meanings, and retaining this as the hypothesis meaning until it is excluded by crossing the threshold for elimination. At this point, the learner then randomly selects a new hypothesis from the remaining (i.e., un-eliminated) candidate meanings. The net effect of this is that at the late times that are relevant to our calculations, the learner will be left with two candidate meanings for a word: the target meaning, and one non-target meaning. They then have a 50\% chance of having hypothesized the target meaning, which results in the threshold $\epsilon$ that determines the lexicon learning time increasing by a factor of $2$. This typically leads to a modest decrease (around $30\%$) in the time to learn a lexicon, which is rather less than what is achieved by a mutual exclusivity constraint, for example. These considerations provide further demonstration \citep[see also][]{Smith2011,Yurovsky2015,Roembke2016} that hypothesis testing and statistical learning aren't mutually exclusive, but are two strategies that can be combined (potentially with others, like mutual exclusivity) to facilitate the acquisition of large lexicons.

A further appealing feature of the model presented here is that it may potentially be more robust to erroneous input than the versions with infinite memory \citep{Smith2006,Blythe2010,Reisenauer2013,Blythe2016}. The key assumption that runs throughout these works (and here) is that the target meaning is always inferred as a candidate meaning by the learner. Various strategies have been discussed for recovering from such errors \citep[e.g.][]{Siskind1996,Tilles2012,Blythe2016}: being able to forget unrepresentative episodes provides another strategy \citep{Ibbotson2018}. Specifically, if the target meaning is missing in one episode, then the association with a non-target meaning may be greater, but not sufficiently large for the threshold to eliminate the target to be reached. For this to happen, the word would need to be exposed several times in quick succession without the target being inferred, the probability of which may be sufficiently small that it rarely occurs in practice. It would be interesting to incorporate this into the present model, and determine its effect on the learning time.

Our findings are, of course, subject to certain caveats. As noted in Section~\ref{sec:moddef}, an exponential decay of associations over time is not well supported empirically. However, as we observed, other proposed models, such as power-laws and decays to a nonzero asymptote, imply superior long-term storage of associations than in the exponential model, and translate to a less stringent limit on the range of reasonable memory lifetimes than was considered in Section~\ref{sec:60k}.

Perhaps the biggest uncertainty arising from our model lies in the process of elimination to determine a meaning for a word, and in particular, the interpretation of threshold $\Gamma$ at which elimination of a candidate meaning takes place.  It is worth recalling that this threshold appeared in the model almost by logical necessity. In Section~\ref{sec:moddef}, we observed that memory decay has no effect on the learning time if only the strongest word-meaning associations are used to identify candidates for a word's meaning. This would be the case if, for example, a learner employed a hypothesis-testing algorithm in which the choice of hypothesis was determined by the maximal association strength for a given word. There are perhaps other viable strategies, for example, where a hypothesis is biased towards the meanings with a higher association; one may then find that the strength of this bias could be related in some way to the elimination threshold that was a key parameter in the present work. A further concern is that all of these strategies implicitly invoke long-term memory: for example, elimination of a meaning or the formation of a hypothesis are typically assumed to have indefinite effect. It might therefore be worthwhile to quantify in a systematic and consistent way the memory burdens that are implied by different models and strategies, so that these may then be compared on an equal footing. Relatedly, the mutual exclusivity constraint modeled here remains highly idealized, with the learner being able to apply this instantaneously across the entire lexicon. It would be useful to characterize all of these processes more precisely from empirical studies, so we may better to understand their impact on the capacity of learners to acquire a human-scale lexicon.

\section*{Acknowledgments}

The authors thank Kenny Smith for many detailed comments on the manuscript.

\appendix

\section{Monte Carlo simulation algorithms}
\label{sec:algos}

Here we set out the Monte Carlo simulation algorithms that were used to obtain the data shown in the main text\footnote{Reference implementations of these algorithms will be archived at a permanent URL on publication.}.

\subsection{Independent word learning}

The model is defined by the number of words $W$, their frequencies $\phi_w$, the number of non-target meanings for each word $M$, their inference probabilities $a_{wm}$, the memory decay rate $\kappa$ and the elimination threshold $\Gamma$.

\begin{enumerate}
\item Set an elapsed time counter $t=0$, and also a previous exposure time $t_w=0$ for each word $w$.
\item For each word-meaning pair, set the relative association strength $x_{wm} \equiv A_{w\target{w}}-A_{wm}=0$.
\item Place all words $w$ in a lexicon ${\cal L}$ of unlearnt words.
\item While ${\cal L}$ contains at least one word, iterate the following steps:
\begin{enumerate}
\item Increase the elapsed time $t$ by an amount $\delta t$ drawn from an exponential distribution with mean
\begin{equation}
\overline{\delta t} = \frac{1}{\sum_{w \in {\cal L}} \phi_w}
\end{equation}
where the sum is over unlearnt words,
\item Sample word $w$ from the set ${\cal L}$ with probability $\phi_w / \overline{\delta t}$, where $\overline{\delta t}$ is given by the above expression.
\item For each meaning $m$ for which $x_{wm}<\Gamma$, iterate the following steps:
\begin{enumerate}
\item Multiply $x_{wm}$ by ${\rm e}^{-\kappa(t - t_w)}$.
\item Increase $x_{wm}$ by $1$.
\item If the result of this increment is that $x_{wm}\ge\Gamma$ for all non-target meanings $m$, the word is now learnt and thus removed from the set ${\cal L}$.
\end{enumerate}
\item Set the previous word exposure time $t_w$ to the current time $t$.
\end{enumerate}
\end{enumerate}

When this algorithm terminates, $t$ contains the learning time for the lexicon, and $t_w$ the time at which each word $w$ was last exposed before becoming learnt.

\subsection{Word learning with a mutual exclusivity constraint}

The algorithm is similar to above, with the following modifications:
\begin{enumerate}
\setcounter{enumi}{3}
\item
\begin{enumerate}
\setcounter{enumii}{2}
\item
\begin{enumerate}
\setcounter{enumiii}{2}
\item If the result of this increment is that $x_{wm}\ge\Gamma$ for all non-target meanings $\Gamma$, the word is now learnt and placed on a queue of recently-learnt words ${\cal Q}$.
\end{enumerate}
\stepcounter{enumii}
\item While ${\cal Q}$ is non-empty:
\begin{enumerate}
\item Pop the word $w$ from the head of ${\cal Q}$, and remove it from the set of unlearnt words ${\cal L}$.
\item For all words $w'$ in ${\cal L}$ iterate the following steps
\begin{enumerate}
\item If $x_{w'\target{w}}<\Gamma$, set it to $\Gamma$. (Recall $\target{w}$ is the target meaning of word $w$).
\item If the result of this increment is that $x_{w'm}\ge\Gamma$ for all non-target meanings $m$, the word $w'$ is now learnt and is pushed onto the queue ${\cal Q}$.
\end{enumerate}
\end{enumerate}
\end{enumerate}
\end{enumerate}

\section{Calculating learning times: independent word learning}
\label{sec:tcalc}

The method for calculating learning times for a lexicon has been developed over a number of earlier works \citep{Smith2006,Blythe2010,Reisenauer2013,Blythe2016}, and different aspects of these calculations are scattered around these various papers. We therefore find it useful in this appendix to set out the entire procedure in a single place for easier reference.

\subsection{Expressions for the lexicon learning probability}

Recall that the learning time $t^\star$ is obtained from the probability $L(t)$ that the entire lexicon has been learnt by a time $t$ via Eq.~(\ref{tstar}). An exact formal expression for the learning probability can be written down by appealing to the quantity
\begin{equation}
C_{wm}(t) = \left\{ \begin{array}{ll} 1 & \mbox{if meaning $m$ is in contention as a candidate for $w$ at time $t$} \\
0 & \mbox{otherwise} \end{array}\right.
\end{equation}
that is defined for each word meaning pair. We work in a paradigm where initially every non-target meaning is entertained as a possible meaning for a given word $w$, and that these non-target meanings are eliminated over time.  In particular, the average of $C_{wm}(t)$ over all sequences of exposures and inferences, denoted $\langle C_{wm}(t) \rangle$ is equal to the probability that a candidate meaning $m$ has not been eliminated as a possible meaning for word $w$ by time $t$.  In the main text, we have generally assumed that this typically decays exponentially at the elimination rate $r_{wm}$, that is
\begin{equation}
\langle C_{wm}(t) \rangle \sim {\rm e}^{-r_{wm} t} \;.
\end{equation}
Sometimes this assumption is exact; in other cases it holds approximately, and has been verified as a valid assumption by comparison with simulation results (see main text).

Now, for each word $w$, the quantity
\begin{equation}
\prod_{m \ne \target{w}} [ 1 - C_{wm}(t) ]
\end{equation}
is equal to $1$ if \emph{only} the target meaning of word $w$ (denoted $\target{w}$) is in contention at time $t$. In this case, the word is deemed learnt. Otherwise this quantity equals $0$.  Consequently, the probability that the entire lexicon has been learnt after time $t$ can be obtained by multiplying this quantity together for each word, and taking an average over all exposures and inferences. That is,
\begin{equation}
L(t) = \left\langle \prod_{w=1}^{W} \prod_{m \ne \target{w}} \left[ 1 - C_{wm}(t) \right] \right\rangle \;.
\end{equation}

This last equation is exact, and becomes easier to handle at late times (i.e., those where the lexicon has a very high probability of being learnt). At these late times, it will be the case that $\langle C_{wm}(t) C_{wm'}(t) \rangle \ll \langle C_{wm}(t) \rangle$ for any pair of meanings $m$ and $m'$. The reason for this is that the probability that two specific meanings have both yet to be eliminated decays more rapidly than the probability that either meaning individually is eliminated. Moreover, the number of exposures of different words, and their contexts of use, are always taken to be completely independent, which means that we have $\langle C_{wm}(t) C_{w'm'}(t) \rangle = \langle C_{wm}(t) \rangle \langle C_{w'm'}(t) \rangle$ for all pairs of words and meanings. Given these approximations, we can write that
\begin{equation}
\label{appLt}
L(t) \sim \prod_{w=1}^{W} \prod_{m \ne \target{w}} \left[ 1 - \langle C_{wm}(t) \rangle \right]  = \prod_{w=1}^{W} \prod_{m\ne \target{w}} \left[ 1 -{\rm e}^{-r_{wm} t} \right] 
\end{equation}
which is Eq.~(\ref{Lt}) in the main text, and the one from which all the learning times are derived.

\subsection{Identifying the slowest decaying terms}
\label{sec:slowdecay}

To convert the expression (\ref{appLt}) into a learning time for the lexicon, we have to identify the slowest decaying terms in the product. In Section~\ref{sec:tlearn} of the main text we describe in detail the case of the uniform word and meaning distributions: here all terms decay at the same rate, and the solution of the equation $L(t^\star)=1-\epsilon$ is straightforwardly obtained.

Different terms decay at different rates when either the word or the meaning frequency distributions are nonuniform. In the case where memory is infinite, and there is no association threshold for excluding a meaning from consideration, the elimination rate $r_{wm} = \phi_w (1-a_{wm})$, that is, the rate at which word $w$ is presented multiplied by the probability that the meaning $m$ is \emph{not} inferred alongside. The slowest decays then typically come from the \emph{lowest} frequency words, but the \emph{highest} frequency (non-target) meaning. This makes sense: the hardest words to learn are those that determine the lexicon learning time; and words are hard to learn if they are not encountered very often, and if there is a high probability of inferring an incorrect meaning.

Throughout this work, we assume power-law (Zipf) distributions for frequencies, as these are attested at least for word forms. These have the property that at the top end of the distribution, the frequencies are well separated, whereas at the bottom end of the distribution, many words (or meanings) have similar frequencies. In the former case, one can simply drop all the slower decays from the product (\ref{appLt}). Therefore, we can truncate the product over non-target meanings to the single most frequent non-target meaning, the consequences of which are described in Section~\ref{sec:tlearn} of the main text. That is, we identify $a$ with the most frequent non-target meaning, and set the number of non-target meanings equal to $1$.

To deal with a non-uniform \emph{word} distribution, we have many terms with similar decay rates (arising from the fact that there are many low-frequency words with similar frequencies). To illustrate how to handle this case, we take a \emph{uniform} meaning distribution combined with a \emph{nonuniform} word distribution. With infinite memory and non elimination threshold, $r_{wm} \equiv r_w = \phi_w (1-a)$, and we have from (\ref{appLt}) that
\begin{equation}
L(t) \sim \prod_{w=1}^{W} \left[ 1 - {\rm e}^{-r_w t} \right]^M \approx 1 - M \sum_{w=1}^{W} {\rm e}^{-r_w t} \;.
\end{equation}
We obtained this approximation by expanding both products, and keeping only terms of the form ${\rm e}^{-r_w t}$ while dropping faster decays like ${\rm e}^{-(r_w+r_{w'})t}$.  The strategy now is to convert the sum to an integral, and to Taylor expand $r_w$ around $w=W$ as $r_w \approx r_W + r'_W (w-W)$, where
\begin{equation}
r'_W = \left. \frac{{\rm d}}{{\rm d} w} r_w \right|_{w=W} \;.
\end{equation}
Note that it is assumed that $w$ is a continuous variable for the purpose of taking the derivative, and that as words are ranked by decreasing frequency, this derivative will be negative. Then we find
\begin{equation}
\label{zipfint}
L(t) \approx 1 - M {\rm e}^{-r_W t} \int_0^W {\rm d} w {\rm e}^{-r'_W (w-W) t} \approx 1 - \frac{M}{|r'_W|t} {\rm e}^{-r_W t} 
\end{equation}
where to obtain the latter approximation we have assumed $r'_W$ is sufficiently large that the lower limit of the integral can be taken down to $-\infty$. Setting this expression for $L(t)$ equal to $1-\epsilon$, as required by Eq.~(\ref{tstar}), we obtain the result (\ref{tzipf}) given in the main text.

If we combine a non-uniform word distribution with a non-uniform meaning distribution, we do the same as before. That is, we set $a$ equal to the frequency of the most frequent meaning, and put the number of non-target meanings $M$ equal to $1$.

\section{Calculating learning times: mutual exclusivity}
\label{sec:meuni}

The general expressions that apply to cross-situational learning with a mutual exclusivity constraint were derived by \cite{Reisenauer2013}. We recapitulate the main points here, and illustrate in particular why a correlation between word and meaning frequencies is required for the mutual exclusivity constraint to have a strong effect.

\subsection{Expressions for the lexicon learning probability}

\cite{Reisenauer2013} identified two conditions that must be met for a lexicon to be learnt when a mutual exclusivity constraint is operating. First, every word in the lexicon must have been exposed to the learner at least once. Second, there must be exactly one way to assign the remaining candidate meanings to words that respects mutual exclusivity (i.e., each word has a unique meaning). In the case where there are meanings that are not the referent of any word in the lexicon there is a further condition; here we restrict ourselves to the case where all meanings that might be encountered by a learner have a corresponding word in the lexicon.

To obtain the analog of (\ref{appLt}), we need to introduce a further set of random variables $E_w(t)$ with the property
\begin{equation}
E_w(t) = \left\{ \begin{array}{ll} 1 & \mbox{if word $w$ has been encountered by time $t$} \\
0 & \mbox{otherwise} \end{array} \right. \;.
\end{equation}
Then, the two conditions given above translate to the expression
\begin{equation}
\label{Ltmecc}
L(t) = \left\langle \left( \prod_{w=1}^{W} E_w\right) \left( \prod_{w_1, w_2, \ldots, w_n} \left[ 1- C_{w_1\,\target{w_2}} C_{w_2\,\target{w_3}} \cdots C_{w_n\,\target{w_1}} \right] \right) \right\rangle 
\end{equation}
where the average is over all sequences of word presentations and inferences drawn by the learner. To lighten the notation, we have dropped the dependence of $E_w$ and $C_{wm}$ on $t$. With this in mind, let us now go through this expression piece by piece. The first term in round brackets is equal to $1$ if and only if all words have been encountered by time $t$. Therefore only sequences of word presentations for which this is true contribute to the lexicon learning probability $L(t)$, as required. The second product is over all subsets of words of size $n \ge 2$. The product over the indicator variables $C$ is equal to $1$ if we have not yet eliminated the target of $w_2$ as a candidate for $w_1$ by cross-situational learning, nor the target of $w_3$ as a candidate for $w_2$ and so on, and furthermore, we have not yet eliminated the target of $w_1$ as a candidate for $w_n$. Under these conditions at least two possible assignments of words to meanings that are compatible with the mutual exclusivity constraint: the target meanings, and the assignment $\target{w_2}$ to $w_1$, $\target{w_3}$ to $w_2$ and so on. Since then some ambiguity remains even under mutual exclusivity, the lexicon cannot be learnt, and the corresponding sequences of exposures and inferences are again prevented from contributing to the probability $L(t)$.

\subsection{Identifying the slowest decaying terms}

As in the case of independent word learning, described in Appendix~\ref{sec:tcalc}, the key to converting (\ref{Ltmecc}) to a learning time is to identify the slowest decaying terms. This time they can come from either the first term (which corresponds to waiting for the least frequent word to be exposed) or the second term (which corresponds to waiting for ambiguity between different assignments of meanings to words to be resolved). 

Now, the first process is equivalent to learning each word in the lexicon when they can all be fast-mapped (i.e., learnt on their first exposure). This would occur if there was a single non-target meaning whose inference frequency $a_{wm}=0$. Hence we find from (\ref{appLt}) that the probability that all word have been exposed (or fast-mapped) by time $t$ is
\begin{equation}
\label{Ltfm}
L_{\rm FM}(t) = \prod_{w=1}^{W} \left[ 1 - {\rm e}^{-\phi_w t} \right]
\end{equation}
where $\phi_w$ is the frequency of word $w$. This equation appears in the main text as Eq.~(\ref{LFM}). Corresponding to this equation is a fast mapping time, $t^\star_{\rm FM}$, given by $L_{\rm FM}(t^\star_{\rm FM}) = 1 -\epsilon$.  For a uniform distribution this is given by (\ref{tuniform}) with $a=0$ and $M=1$; and for a Zipf distribution by (\ref{tzipf}) with $a=0$ and $M=1$.

We now turn to the time associated with resolving ambiguity between different assignments of meanings to words. The general rule is that the longer the subset of words $w_1, w_2, \ldots, w_n$ in (\ref{Ltmecc}) the faster the decay. Thus, as described in the main text, the long-time behaviour of the learning probability can be obtained by keeping only the shortest subsets, that is, pairs of words.  This leads to the approximation
\begin{equation}
\label{Ltpairs}
L_{\rm pairs}(t) \approx \left\langle \prod_{\langle w, w'\rangle} \left[1 - C_{w\,\target{w'}} C_{w'\,\target{w}} \right] \right\rangle \approx \prod_{\langle w, w'\rangle} \left[ 1 - {\rm e}^{-(r_{w\,\target{w'}}+r_{w\,\target{w'}})t} \right]
\end{equation}
to the second term in (\ref{Ltmecc}). Associated with this is a pair resolution time, $t^\star_{\rm pairs}$, given by $L_{\rm pairs}(t^\star_{\rm pairs}) = 1 - \epsilon$.

To determine the lexicon learning time under mutual exclusivity, the procedure is to calculate each of the two times $t^\star_{\rm FM}$ and $t^\star_{\rm pairs}$. The longer of these two times then gives the lexicon learning time. This procedure is equivalent to solving $L_{\rm ME}(t)=1-\epsilon$ given by (\ref{Ltme}) in the main text. In the following we illustrate with specific examples.

\subsubsection{Uniform word and meaning distribution}

\begin{figure}
\begin{center}
\includegraphics[width=0.66\linewidth]{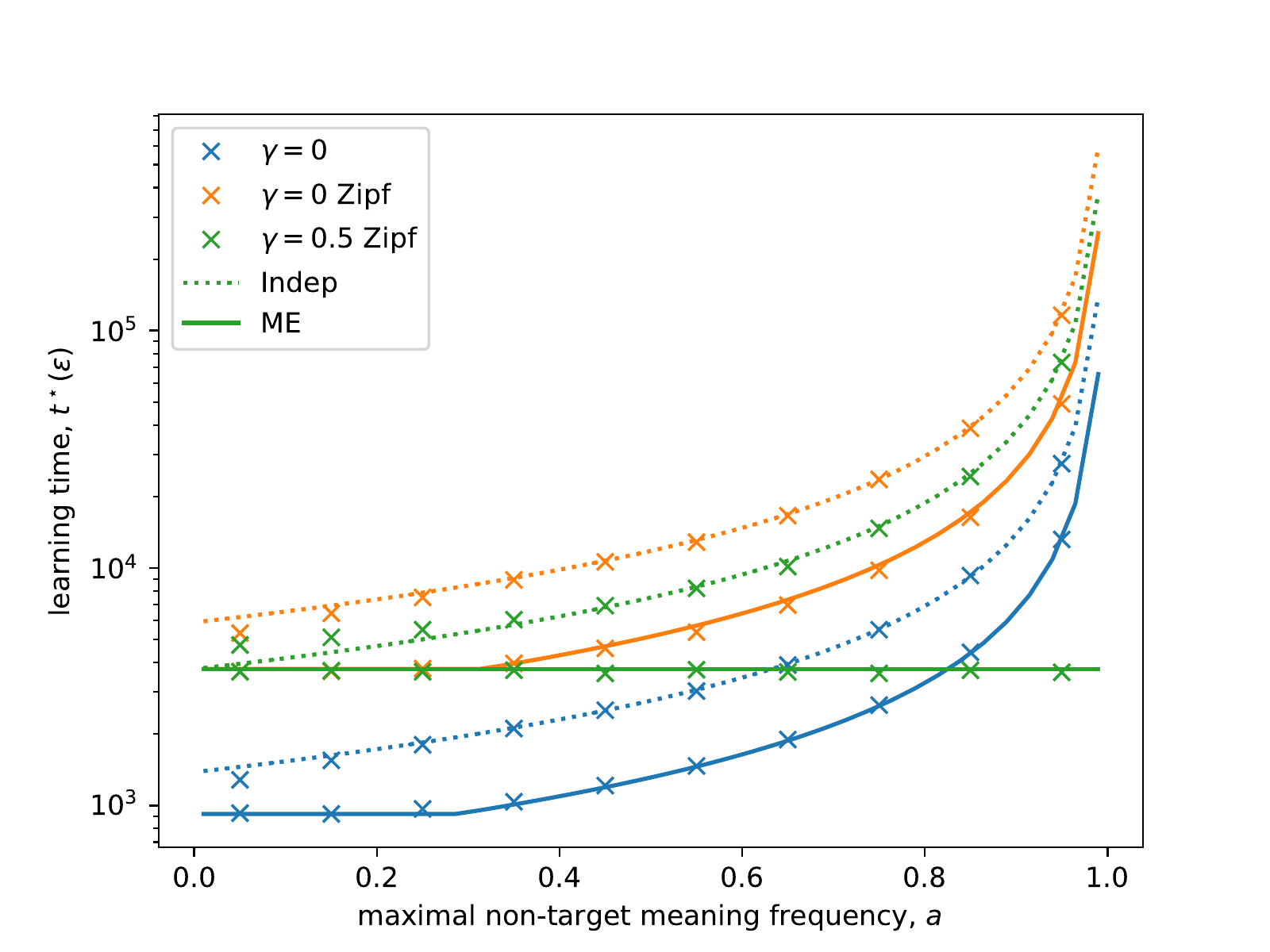}
\end{center}
\caption{\label{fig:w100m99} Learning time for a lexicon of $W=100$ words with (solid lines) and without (dotted lines) a mutual exclusivity constraint under different word and meaning distributions. The meaning frequency distribution is of the form $a_{wm} = a/m^{\gamma}$, with $\gamma=0$ indicating a uniform distribution, and $\gamma=\frac{1}{2}$ the inverse square-root form discussed in the main text. A Zipf word distribution tends to increase the learning time, while a non-uniform meaning distribution tends to decrease it. The mutual exclusivity constraint has the biggest impact when meaning and word frequency distributions are nonuniform and correlated.  Lines are theoretical predictions, crosses are obtained from Monte Carlo simulation data.}
\end{figure}

In the first set of examples, we consider for simplicity infinite memory and no threshold for elimination of a non-target meaning. With both a uniform word and meaning distribution we have $r_{wm} = \phi_w(1-a_{wm})$ as previously, and $\phi_w=1/W$ and $a_{wm}=a$.  From (\ref{Ltfm}) we find
\begin{equation}
t_{\rm FM}^\star = - W \ln\left[ 1 - (1-\epsilon)^{\frac{1}{W}}\right]
\end{equation}
and from (\ref{Ltpairs}) that
\begin{equation}
\label{tmeuniform}
L_{\rm pairs}(t) \approx \left[1 - {\rm e}^{-2(1-a)t/W} \right]^{W(W-1)/2} \implies
t^\star_{\rm pairs} = - \frac{W \ln[1 - (1-\epsilon)^{\frac{2}{W(W-1)}}]}{2(1-a)}
\end{equation}
where we have used the fact that $M=W-1$.  When the inference probability $a$ is sufficiently small, it can be the case that $t_{\rm FM}^\star < t^\star_{\rm pairs}$, and the low-frequency words in the lexicon are effectively fast-mapped (because the non-target meanings that appear in their context have been eliminated via the mutual exclusivity constraint). In Fig.~\ref{fig:w100m99} we plot the learning time (solid blue curve for the case of uniform word and meaning frequencies), and see that there is a kink at around $a=0.3$ which is the point at which resolving ambiguity between pairs of words starts to limit the rate at which the lexicon can be acquired.

The figure also shows the case where words are learnt independently (dotted curves). For the case of the uniform word and meaning distributions, the result (\ref{tuniform}) with $r=(1-a)/W$ and $M=W-1$ applies, that is,
\begin{equation}
t^\star_{\rm ind} = - \frac{W  \ln\left[ 1 - (1-\epsilon)^{\frac{1}{W(W-1)}} \right]}{1-a} \;.
\end{equation}
This is in fact somewhat similar in form to (\ref{tmeuniform}). The factor of $2$ in the denominator in (\ref{tmeuniform}) indicates that mutual exclusivity roughly halves the learning time when words and meanings are uniformly distribution (the other factor of $2$ under the logarithm has a more modest effect on the learning time).

\subsubsection{Zipfian word and uniform meaning distribution}

We now consider what happens if the word distribution has a Zipf distribution, while retaining a uniform distribution of meanings.  Recall from Section~\ref{sec:slowdecay} that there are many words at the bottom end of the distribution, which means that the sums obtained by expanding the products (\ref{Ltfm}) and (\ref{Ltpairs}) need to be replaced with integrals.  For the case of the former, this has already been done: we obtain the fast mapping time by putting $a=0$ and $M=1$ in (\ref{tzipf}). This yields
\begin{equation}
\label{t0zipf}
t^\star_{\rm FM} = WZ \Lambert \left( \frac{W}{\epsilon} \right) \;,
\end{equation}
where $Z$ is the normalization of the Zipf distribution, given by (\ref{zipf}), and $\Lambert(z)$ is the principal branch of the Lambert W function.

Meanwhile, expanding (\ref{Ltpairs}) with $M=W-1$, we obtain
\begin{equation}
\label{Ltmeint}
L_{\rm pairs}(t) \approx 1 - \frac{1}{2} \int_0^W {\rm d} w \int_0^{W-1} {\rm d} m \, {\rm e}^{-(r_{wm}+r_{mw})t} \approx 1 - \frac{{\rm e}^{-2 \tilde{r} t}}{2 (\tilde{r}')^2 t^2 }
\end{equation}
where 
\begin{align}
\label{tilder}
\tilde{r} &= r_{W,W-1} \\
\label{tilderp}
\tilde{r}' &= \left. \frac{{\rm d}}{{\rm d} w} (r_{wm}+r_{mw}) \right|_{w=W, m=W-1} \;.
\end{align}
The solution to the equation $L_{\rm pairs}(t^\star_{\rm pairs}) = 1-\epsilon$ is given in the main text as Eq.~(\ref{tastme}).

In the case of infinite memory and no threshold, $r_{wm} = \phi_w(1-a_{wm})$. Taking also the specific choices of the Zipf distribution (\ref{zipf}) for word forms, and a uniform meaning distribution with $M=W-1$ non-target meanings, we have
\begin{equation}
\tilde{r} = \frac{1-a}{WZ} \quad\mbox{and}\quad \tilde{r}' = - \frac{1-a}{W^2 Z}  \;.
\end{equation}
Substituting these into (\ref{tastme}) we obtain
\begin{equation}
\label{tmezipf}
t^\star_{\rm pairs} = \frac{W Z}{1-a} \Lambert \left(  \frac{W}{\sqrt{2\epsilon}} \right) \;.
\end{equation}

Again it is the longer of $t^\star_{\rm FM}$ and $t^\star_{\rm pairs}$ that determines the lexicon learning time. This is plotted as a function of the degree of referential uncertainty, $a$, in Fig.~\ref{fig:w100m99} as the orange curves. One sees that the learning time is systematically higher than when words are uniformly distributed.

To understand this slowdown in more detail, we note that when $W$ is large or $\epsilon$ small (as is usually assumed), the Lambert W function can be approximated by the natural logarithm. This makes comparison of different conditions easier. First, we find that since the argument of the Lambert W function in (\ref{tmezipf}) is essentially the square root of that in (\ref{tzipf}), mutual exclusivity again serves to roughly halve the learning time when the meaning distribution is uniform, independently of the word distribution. Second, when we compare with the case of the uniform word distribution, we find overall that the effect of Zipf distributed words is to multiply the learning time by a factor of $Z$, defined in Eq.~(\ref{zipf}), independently of whether the mutual exclusivity constraint is operating or not. These observations all assume that the inference probability $a$ is sufficiently high that the resolution of ambiguous pairs of words determines the learning time. Although mutual exclusivity does accelerate word learning, when the meaning distribution is uniform, it is (except for very small lexicons) insufficient to counteract the slowdown that arises from a Zipf word distribution.

\subsubsection{Zipf word distribution and nonuniform meaning distribution}

In the main text, we restricted our discussion to the case where the word have a Zipf distribution, and the meaning distribution was of the inverse square-root form (\ref{invsqrtme}). We also assumed that high-frequency meanings were the targets of high-frequency words. As we now show, this allows the mutual exclusivity constraint to deliver rapid word learning.

In this case, the fast mapping time continues to be given by (\ref{t0zipf}), as this depends only on the word form distribution. Meanwhile, \citet{Reisenauer2013} showed that the two most frequent words are the slowest to disambiguate. That is, in (\ref{Ltpairs}), we simply keep the term with $w=1, w'=2$.  Then, we find that
\begin{equation}
L_{\rm pairs}(t) \approx 1 - {\rm e}^{-(\phi_1+\phi_2)(1-a) t} \implies t_{\rm pairs}^\star = - \frac{\ln\epsilon}{(\phi_1+\phi_2)(1-a)} = - \frac{2Z \ln\epsilon}{3(1-a)} \;.
\end{equation}
Once again, it is the longer of these two times that determines the lexicon learning time.

A crucial feature of this result is that it increases with the lexicon size $W$ only through the quantity $Z$ that normalizes the Zipf distribution (which grows roughly logarithmically with $W$). Compared to the case of a Zipf word distribution combined with a uniform meaning distribution, the rate of word learning is of order $W \ln W$ faster, at least in the regime where resolution of ambiguity between pairs of words determines the learning time.  In other words, the correlation between word frequencies and the frequencies that their correspond meanings are inferred as candidates for other words can be exploited by the mutual exclusivity constraint to deliver rapid learning. As can be seen from Fig.~\ref{fig:w100m99} (green curves) when the non-target meaning frequency $a$ is sufficiently large, it is possible for the slowdown that occurs as a result of the Zipf word distribution to be counteracted by the gain arising from the mutual exclusivity constraint.

\subsubsection{Extension to the case with a finite memory}

As with independent learning, the results set out above can be generalized to the case of a finite memory by replacing the expressions for the elimination rates $r_{wm}$ with the appropriate result, Eq.~(\ref{r1}).  Memory decay is irrelevant to the fast mapping times, $t^\star_{\rm FM}$, and so these do not need to be recalculated. Moreover, with memory decay operating we have not encountered any situations where this exceeds the time to resolve ambiguity between pairs of words. As explained in the main text, this is due to an important difference with the case of infinite memory: the slowest decaying pairs now correspond to the \emph{least frequent} words. 

We obtain the pair resolution time $t^\star_{\rm pairs}$ from (\ref{Ltmeint}) with the elimination rate $r_{wm}$ appropriate redefined to incorporate the effect of memory decay, that is, with
\begin{equation}
r_{wm} = \phi_w (1-a_{wm}) \lambda\left( \frac{\kappa}{\phi_w(1-a_{wm})} \right) 
\end{equation}
which follows from (\ref{r1}).  For the specific case of the Zipf word distribution and the inverse square-root meaning distribution, we  obtain from (\ref{tilder}) and (\ref{tilderp}) the expressions (\ref{tastme})--(\ref{endtastme}) given in the main text.

\section{Calculating learning times: infinite memory with a threshold}
\label{sec:kappa0}

As noted in the main text, the case of infinite memory, $\kappa=0$, requires a special treatment. The case $\Gamma<1$ was treated in earlier works \citep{Smith2006,Blythe2010,Reisenauer2013}; however the case $\Gamma>1$ is different because a non-target meaning has to be absent more than once in order to be eliminated, even if memories don't decay. Specifically, the word $w$ needs to be presented, and the non-target meaning $m$ needs to be absent $\lceil \Gamma \rceil$ times (where $\lceil \Gamma \rceil$ is the smallest integer that is equal to, or larger than, $\Gamma$) in order for the elimination threshold $\Gamma$ to be crossed.

If the word $w$ is presented as a Poisson process with rate $\phi_w$, and the probability that on each of these occasions the meaning $m$ is inferred is $a_{wm}$, the association between word $w$ and meaning $m$ jumps one unit closer to the threshold at $\Gamma$ as a Poisson process with rate $r_{wm} = \phi_w(1-a_{wm})$. Recalling that $C_{wm}(t)=1$ until this threshold is crossed, and $0$ thereafter, we have in the case of infinite memory that
\begin{equation}
\label{Cwminf}
\langle C_{wm}(t) \rangle = \sum_{n=0}^{\lceil \Gamma \rceil-1} \frac{(r_{wm} t)^n}{n!} {\rm e}^{-r_{wm} t} \;,
\end{equation}
that is, the probability that fewer than $\lceil \Gamma \rceil$ steps towards the threshold have been taken by time $t$. It is the fact that this decay is not purely exponential that makes this case special.

To make use of this expression, it now has to be substituted into Eq.~(\ref{appLt}) which gives the learning probability $L(t)$ for the case of independent learning, or its analog Eq.~(\ref{Ltme}) for the case of learning with a mutual exclusivity constraint. The learning time $t^\star$ is obtained, as before, by putting $L(t^\star) = 1-\epsilon$. In general, this does not have a closed-form expression, even in terms of special functions, and so we solve these cases with a root-finding algorithm (specifically, the SciPy routine \texttt{fsolve}, \citealp{Scipy}). 

For example, in the case of independent learning and a uniform word and meaning distribution, we solve
\begin{equation}
\sum_{n=0}^{\lceil \Gamma \rceil-1} \frac{(r t)^n}{n!} {\rm e}^{-r t}  = 1 - \left[1 - (1-\epsilon)^\frac{1}{MW} \right]
\end{equation}
to find the learning time [cf., Eq.~(\ref{tuniform})], where $r=\phi(1-a)$, $\phi=1/W$ is the uniform word frequency and $a$ is the probability that each of the $M$ non-target meanings is inferred at each exposure. As previously, we can put $M=1$ and take $a$ equal to the maximal non-target meaning frequency to handle the case of a nonuniform meaning distribution. 

To handle the case of a Zipf word distribution, we treat the polynomial prefactor of the exponential decay in (\ref{Cwminf}) as constant while integrating over the bottom end of the word distribution. (This is based on the fact that the polynomial varies less rapidly than the exponential factor). This means that Eq.~(\ref{zipfint}) for the learning probability becomes
\begin{equation}
L(t) \approx 1 - \frac{M}{|r'_W|t} \sum_{n=0}^{\lceil \Gamma \rceil-1} \frac{(r_W t)^n}{n!} {\rm e}^{-r_W t}
\end{equation}
where $r_W$ and $r'_W$ have the same meanings as in Section~\ref{sec:tlearn}. We set this equal to $1-\epsilon$ and solve numerically for $t^\star$. Again, a nonuniform meaning distribution can be handled by approximating it by the single most frequent meaning.

The same strategy is used in the case of the integral that is involved in the calculation of the learning probability under the mutual exclusivity constraint (\ref{Ltmeint}). In this case we arrive at the expression
\begin{equation}
L_{\rm pairs}(t) \approx 1 - \frac{{\rm e}^{-2\tilde{r}t}}{2(\tilde{r}')^2t^2} \sum_{n=0}^{\lceil \Gamma \rceil-1} \frac{(\tilde{r}t)^n}{n!} 
\end{equation}
where here
\begin{align}
\tilde{r} &= \frac{1}{WZ}\left(1 - \frac{a}{\sqrt{W}}\right) \\
\tilde{r}' &= - \frac{1}{W^2Z} \left[ 1 - \frac{3a}{2\sqrt{W}} \right] \;.
\end{align}
Again, we set this expression for $L(t)=1-\epsilon$ to obtain the learning time $t^\star$.

\section{Elimination rate for large decay rates or thresholds}
\label{sec:asymp}


In this final appendix we derive an analytical formula for the smallest eigenvalue of (\ref{eigen}) that is valid in the limit of a large decay rate $\kappa$ or threshold $\Gamma$, i.e., the regime where words become hard to learn and the learning time is correspondingly large. In the main text, we found this to agree well with simulation results even for fairly modest decay rates and thresholds, and it may therefore be preferred to direct numerical solution of Eq.~(\ref{eigen}).

We first state the result, and then show how it is obtained. In the large $\kappa$, large $\Gamma$ regime, the smallest eigenvalue can be approximated as
\begin{equation}
\label{lasymp}
\lambda(k) = k\Gamma \sqrt{\frac{k|s|}{2\pi|k\Gamma(s+1)-1|}} \exp\left[ s\Gamma - \frac{\Ein(s)}{k} \right]
\end{equation}
where 
\begin{equation}
\label{Ein}
\Ein(z) = \int_0^z {\rm d}u \frac{1-{\rm e}^{-u}}{u}
\end{equation}
and $s$ is the solution of the equation
\begin{equation}
\label{simplicit}
k \Gamma s = 1 - {\rm e}^{-s}
\end{equation}
that tends to $-\infty$ as $\Gamma\to\infty$.  The simplification (\ref{bigbig}) is obtained for very large values of $\kappa$ and $\Gamma$, since the solution of (\ref{simplicit}) approaches $s \sim -\ln(k\Gamma)$ and $\Ein(s) \sim  - \frac{k\Gamma}{\ln(k\Gamma)}$. We obtained Eq.~(\ref{bigbig}) by appealing to (\ref{lasymp}), (\ref{r1}) and (\ref{tstar1}).


The derivation of these results begins by taking the continuum limit of (\ref{eigen}), i.e., the limit $N\to\infty$ with $x = n/N$ constant. The resulting equation takes the form of a delay differential equation
\begin{equation}
\label{eigenct}
 \phi(x-1) - \phi(x) + k \frac{{\rm d}}{{\rm d} x} x \phi(x) = -\lambda(k) \phi(x) \;.
\end{equation}
Given the form of the eigenfunction $\phi(x)$, we can integrate this equation on both sides from $0$ to the boundary value $x=\Gamma$ to obtain
\begin{equation}
\label{lamintb}
\lambda(k) = \frac{\int_{\Gamma-1}^\Gamma {\rm d} x \phi(x) - k  \Gamma \phi(\Gamma) }{\int_0^\Gamma {\rm d} x \phi(x) } \;.
\end{equation}
Now we assume that the eigenvalue is sufficiently small that $\lambda(k)$ in (\ref{eigenct}) can be approximated as zero, and that we can approximate $\phi(x)$ through the stationary solution
\begin{equation}
 \phi(x-1) - \phi(x) + k \frac{{\rm d}}{{\rm d} x} x \phi(x) \approx 0
\end{equation}
that is subject to a normalization condition
\begin{equation}
\label{norm}
\int_0^\infty {\rm d} x \phi(x) = 1 \;.
\end{equation}
We now also assume that $k$ or $\Gamma$ are sufficiently large that the stationary solution satisfies $\phi(x)\approx0$ for all $x\ge \Gamma$. Then (\ref{lamintb}) simplifies to
\begin{equation}
\label{lamint}
\lambda(k) \approx \frac{\int_{\Gamma-1}^\Gamma {\rm d} x \phi(x)}{1- \int_{\Gamma}^{\infty} {\rm d} x \phi(x) }  \approx \int_{\Gamma-1}^\Gamma {\rm d} x \phi(x)  \;.
\end{equation}

We solve (\ref{eigenct}) by taking a Laplace transform
\begin{equation}
\tilde{\phi}(s) = \int_0^\infty {\rm d} x {\rm e}^{-sx} \phi(x) \;.
\end{equation}
The transformed version of (\ref{eigenct}) reads
\begin{equation}
\label{eigenlt}
{\rm e}^{-s} \tilde\phi(s) - \tilde\phi(s) - ks \frac{{\rm d}}{{\rm d} s} \tilde\phi(s) = 0
\end{equation}
where we have used the fact that $\phi(x)=0$ for $x<0$ and that, for the eigenfunction to be normalisable, $x\phi(x) \to 0$ as $x\to\infty$. The normalization of $\phi(x)$ also implies the boundary condition $\tilde\phi(0)=1$. The solution of (\ref{eigenlt}) that satisfies this condition is
\begin{equation}
\tilde\phi(s) = \exp\left[ - \frac{1}{k} \Ein(s) \right]
\end{equation}
where the $\Ein$ function is defined by Eq.~(\ref{Ein}).

To invert the Laplace transform for large $x$, we need to apply the saddle-point method \citep{Arfken2012} to the inversion integral
\begin{equation}
\label{brom}
\phi(x) = \frac{1}{2\pi i} \int_{s_0-i\infty}^{s_0+i\infty} {\rm d} s {\rm e}^{sx + \ln \tilde\phi(s)} \end{equation}
where the contour passes through the saddle point $s_0$ along the path of steepest descent (which in this case turns out to be parallel to the imaginary axis). The location of the saddle point is given by the solution of
\begin{equation}
\frac{{\rm d}}{{\rm d} s} \left[sx + \ln \tilde\phi(s)\right] = x - \frac{1}{k} \frac{1-{\rm e}^{-s}}{s} = 0 \;,
\end{equation}
in which the derivative of the $\Ein$ function follows from its definition as an integral, Eq.~(\ref{Ein}). This implies that $s_0$ is a function of $x$, given implicitly by
\begin{equation}
\label{saddle}
k x s_0 = 1-{\rm e}^{-s_0} \;.
\end{equation}
This equation always has a solution at $s_0=0$ and for some $s_0<0$. It is the latter solution that locates the saddle point.  Taylor expanding $sx + \ln\tilde\phi(s)$ to second order about $s=s_0$ yields a Gaussian integral and therewith the large-$x$ expression 
\begin{equation}
\phi(x) \sim \sqrt\frac{k |s_0|}{2\pi | k x (s_0+1) - 1 |} \exp\left[ s_0 x - \frac{1}{k} \Ein(s_0) \right]  \;.
\end{equation}

Estimating the eigenvalue from (\ref{lamint}) requires us to integrate $\phi(x)$ from $\Gamma-1$ to $\Gamma$. Over this region, all contributions to $\phi(x)$ except the exponential part ${\rm e}^{s_0 x}$ vary slowly with $x$. Therefore, we approximate these as constant, and integrate the exponential part:
\begin{equation}
\lambda(k) \approx \sqrt\frac{k |s|}{2\pi | k \Gamma (s+1) - 1 |} \exp\left[ s \Gamma - \frac{1}{k} \Ein(s) \right] \frac{1 - {\rm e}^{-s}}{s}
\end{equation}
where $s$ is given by (\ref{saddle}) with $x=\Gamma$, that is, the solution of Eq.~(\ref{simplicit}). Using (\ref{saddle}), we can rewrite the final term in this expression for the eigenvalue to arrive at the form (\ref{lasymp}).


\end{document}